\documentclass{jfm}
\usepackage{graphicx}
\usepackage{epstopdf, epsfig}
\graphicspath{ {./figures/} }
\usepackage{color}
\definecolor{gray}{rgb}{0.5,0.5,0.5}
\usepackage{psfrag}
\usepackage{amsmath}
\usepackage{mathabx}
\usepackage{tikz}
\usetikzlibrary{shapes}
\usepackage{tabularx}
\usepackage{mathalfa}
\usepackage{multirow}
\usepackage{mathrsfs}
\def\spacce#1{\hskip #1pt}
\def\drawline#1#2{\raise 2.5pt\vbox{\hrule width #1pt height #2pt}}
\def\solid{\drawline{24}{1.0}\nobreak}

\def\bdash{\hbox{\drawline{5.8}{1.0}\spacce{2}}}

\def\dashed{\bdash\bdash\bdash\nobreak}
\def\dotdashed{\bdash\spacce{3}\bdot\spacce{3}\bdash}

\def\bdot{\hbox{\drawline{1}{.5}\spacce{2}}}

\def\trian{\raise 1.25pt\hbox{$\scriptstyle\triangle$}\nobreak}

\def\dtrian{\raise 1.25pt\hbox%
{$\scriptscriptstyle\bigtriangledown$}\nobreak}

\def\squar{\raise 1.25pt\hbox{$\scriptstyle\Box$}\nobreak}

\def\diamon{\raise 1.25pt\hbox{$\scriptstyle\diamond$}\nobreak}

\newcommand{\com}[1]{\textcolor{blue}{#1}}

\def\linedtri1{\hbox{\bdash\hspace{-1.6mm}$\bigtriangleup$\hspace{-0.8mm}\bdash}\nobreak}
\def\soliddtrian1{$\blacktriangledown$\nobreak}
\def\solidrtrian2{$\blacktriangleright$\nobreak}
\def\solidltrian3{$\blacktriangleleft$\nobreak}
\def\beq{\begin{equation}}
\def\eeq{\end{equation}}

\def\We{{\mathrm{\it We}}}
\def\bs{\boldsymbol}

\def\bs{\boldsymbol}

\def\ggg{{\it g}}

\newcommand{\linescircle}[1]{\raisebox{0.5pt}{\tikz{\draw[-,#1!40!#1,solid,line width = 1pt](0,0) -- (8mm,0);\node[draw,scale=0.5,circle,#1!40!#1,fill=white!20!white,line width = 0.9pt]() at (4mm,0.0mm) {};}}}

\newcommand{\linesquare}[1]{\raisebox{0.5pt}{\tikz{\draw[-,#1!40!#1,solid,line width = 1pt](0,0) -- (8mm,0);\node[draw,scale=0.5,regular polygon, regular polygon sides=4,#1!40!#1,fill=white!20!white,line width = 0.9pt]() at (4mm,0.0mm) {};}}}

\DeclareTextFontCommand{\textarial}{}
\definecolor{orange}{RGB}{255,127,0}

\shorttitle{Local and non-local drop deformation in turbulence}
\shortauthor{Alberto Vela-Mart\'in and Marc Avila}

\title{Deformation of drops by outer eddies in turbulence}

\author{Alberto Vela-Mart\'in\aff{1,2}
  \corresp{\email{albertovelam@gmail.com}},
  Marc Avila \aff{1}}

\affiliation{$^{1}$Centre of Applied Space Technology and Microgravity (ZARM) University of Bremen, 28359 Bremen \\
 $^{2}$School of Aeronautics, Universidad Polit\'ecnica de Madrid, 28040 Madrid, Spain}

\begin{document}

\maketitle

\begin{abstract}
Drop deformation in fluid flows is investigated here as an exchange between the kinetic energy of the fluid and the surface energy of the drop. 
We show analytically that this energetic exchange is controlled only by 
the stretching (or compression) of the drop surface by the rate-of-strain tensor. This mechanism is analogous to the stretching of the vorticity field in turbulence. 
Leveraging the non-local nature of turbulence dynamics, 
we introduce a new decomposition that isolates 
the energetic exchange due to local drop-induced surface effects, 
from the non-local action of turbulent fluctuations.
We perform direct numerical simulations 
of single inertial drops in isotropic turbulence 
and show that an important contribution to the increments of the surface energy arises 
from the non-local stretching of the fluid-fluid interface 
by eddies far from the drop surface (outer eddies).
We report that this mechanism is dominant and independent of surface dynamics 
in a range of Weber numbers in which drop breakup occurs.
These findings shed new light on drop deformation and breakup in turbulent flows,
and open the venue for the improvement and simplification of breakup models. 

%The dynamics inside the drop and close to the drop surfac for sufficiently large Weber numbers. 

%These results provide a better understanding of the phenomenlogy of drop breakup and simplifications breakup models.

%Results are discussed in the context of breakup modelling.
\end{abstract}
\begin{keywords}
turbulence, drop breakup, diffuse-interface methods
\end{keywords}

% ------------------------------------------------------------------------------------------
\section{Introduction}
Turbulent binary mixtures of inmiscible fluids are ubiquitous in natural phenomena
and industrial applications. 
Their physical properties depend strongly 
on the structure of the disperse phase, 
in which the breakup of fluid particles---drops or bubbles---plays a fundamental role.
Understanding and modelling turbulent breakup is essential to predict and control 
the dynamics of inmiscible mixtures, but this remains a challenge to theoretical and empirical approaches. 
The physical mechanisms that drive particle deformation and breakup are still poorly understood.

% A fundamental understanding of the mechanisms that drive the deformation and breakup of fluid particles
% is thus essential to advance in 
% is essential to develop accurate breakup models, 

The first studies of drop and bubble breakup in turbulent flows date back 
to the pioneering work of \citet{kolmogorov1949} and \cite{hinze1955},
who proposed the idea of the maximum stable diameter based on dimensional analysis. 
This quantity provides a reasonable estimate of 
the predominant fluid-particle size in turbulent mixtures, 
and has been extensively validated in experiments and numerical simulations \citep{hinze1955,perlekar2012droplet,mukherjee2019droplet,yi2021global}.
Although highly successful, the maximum stable diameter does not provide information on the dynamics of breakup,
which is essential to predict the spatio-temporal variations of fluid-particle distributions in complex flows \citep{hakansson2019}.
% Breakup 
%%%
% predict the temporal evolution of drop populations, 
% a complex problem that remains unsolved, 
% partly because of its inherent  and partly because 
% of our limited understanding of drop dynamics in turbulence.
%%%
The evolution of fluid-particle distributions
are usually described with population balance equations (PBE),
which use convolution kernels to model the effect of breakup and coalescence  \citep[for a recent review see][]{ramkrishna2014population},.
An important parameter of these kernels is the breakup rate,
which depends on the particle size, the material properties of the mixture, and the 
characteristics of turbulence.
A profusion of models predict the breakup rate based on different phenomenological assumptions \citep{lasheras2002review,liao2009literature},
but many of these assumptions are heuristic or cannot be 
readily tested against empirical data.
% due to their ambiguous connection to any quantitative framework.
As a consequence, these models provide contradictory results,
and their ability to yield accurate predictions of 
particle-size distributions even in simple turbulent flows is limited \citep{aiyer2019population}.
In addition, determining the breakup rate directly from experiments is not straightforward \citep{haakansson2020validity},
and data to validate and parameterise these models have been traditionally difficult to obtain.
Laboratory experiments are challenging 
\com{because of the difficulty to isolate surface tension effects from other effects such as buoyancy \citep{risso1998}, 
or large-scale anisotropies and inhomogeneities  \citep{eastwood2004,andersson2006,solsvik2015single},  
or to follow the evolution of single drops until breakup \citep{maass2012}.}
Direct numerical simulations have recently filled this gap by producing high-quality data to study particle-turbulence 
interactions \citep{dodd2016interaction} and particle breakup in turbulent emulsions \citep{perlekar2012droplet,komrakova2015numerical,scarbolo2015coalescence,roccon2017viscosity,mukherjee2019droplet},
and in single drop simulations \citep{qian2006simulation,shao2018,zhong2019liquid}.
A recent review on direct numerical simulations of particle-laden flows is \citet{elghobashi2019}.
Despite these advances, our limited understanding of the theoretical aspects of particle-turbulence interactions 
limits the exploitation of these results for breakup modelling.

Some commonly used breakup models rely on an energetic description of particle deformation and breakup \citep{coulaloglou1976drop,narsimhan1979model,luo1996}.
In this context, the surface energy of the particle is a marker of its degree of deformation,
which increases due to interactions with turbulent fluctuations. 
This process is viewed as an energetic exchange between the turbulent kinetic energy of the flow and the surface energy of the particle.
Similarly to the idea of the activation energy in chemistry, breakup is thought to occur when a threshold of the surface energy is reached \citep{andersson2006}.
A widespread phenomenological picture describes the increments of the surface energy as caused by the 
`impact' of eddies on the surface of the particle \citep[see e.g.][for reviews]{lasheras2002review,liao2009literature}. 
The kinetic energy of these eddies and their arrival frequency are relevant model parameters, and an efficiency factor is usually introduced to account for an incomplete energetic exchange in the process.
The energetic approach to modelling breakup is convenient because it lumps the complexity of particle-turbulence interactions into model parameters,
but it depends on the validity of the model assumptions about the energetic exchange between the surface energy and the kinetic energy of turbulent fluctuations. 

An important limitation of the energetic approach is the lack of a quantitative framework to test and refine these phenomenological breakup models.
In particular, the idea of `impact' and the definition of eddies are fraught with ambiguity,
and the connection of these concepts with measurable quantities in the flow is unclear. 
Recent studies have addressed the problem of the energetic exchange in fluid-fluid dispersions \citep{dodd2016interaction,rosti2019droplets,perlekar2019kinetic,mukherjee2019droplet}, 
but focusing mainly on global budgets, which limits their application to fluid-particle breakup.

This paper addresses the energetic problem of particle-turbulence interactions from a local perspective,
and presents an analytical derivation of the local mechanism responsible 
for the energetic exchange between surface energy and kinetic energy.
For our derivation, we use the Cahn--Hilliard--Navier--Stokes equations \citep{jacqmin1999calculation},
which provide a thermodynamically consistent and convenient framework to study the interactions between turbulence and fluid-fluid interfaces from an energetic perspective.
In these equations the interface has finite thickness, 
but in the limit of a sharp interface the classical stress balance at the infinitesimal fluid-fluid interface is recovered \citep{magaletti2013sharp}.
% The finite thickness of the interface allows to identify  the interactions between the interface and the velocity field.
We exploit this to show that the energetic exchange is solely described by the action of the rate-of-strain tensor on the surface of the fluid particle.
The increments of the surface energy are a consequence of the stretching of the surface by the rate-of-strain tensor, 
a mechanism that resembles the stretching of the vorticity field in turbulence. 

This result applies in general to fluid-fluid interfaces.
Here we use it to study drop deformation in turbulence. 
We analyse data generated by direct numerical simulations of single drops embedded in homogeneous isotropic turbulence in range of Weber numbers in which breakup occurs.
Using analytical tools borrowed from turbulence research \citep{ohkitani1995nonlocal,hamlington2008local},
we distinguish the self-induced increments due to surface (inner) dynamics 
from the action of the surrounding (outer) turbulence.
We show that an important contribution to drop deformation and breakup stems from the non-local surface stretching by eddies away from the drop, 
which constitutes a quantitative reinterpretation of the phenomenological collision of eddies.
This analysis provides a consistent quantitative framework 
to advance the theoretical understanding of breakup and its modelling.

% to advance in the theoretical understanding of breakup and its modelling.

% On the other hand, we show that the local surface stretching is associated with

% tune, improve and develop breakup models. 

%In this paper, we leverage this approach to shed light on the mechanisms of drop deformation and breakup in homogeneous isotropic turbulence. 

% These results pave the way for a better understanding of breakup from an energetic perspective, and provide a consistent quantitative framework to tune, improve and develop breakup models. 

This paper is organised as follows. 
In $\S$\ref{sec:11}, we present the derivation of the fundamental mechanisms that drives the exchange between the kinetic energy and the surface energy. We study this mechanism in direct numerical simulations of single drops, which are described in $\S$\ref{sub:nm}. The results and their discussion are presented in $\S$\ref{sec:anal} and $\S$\ref{sec:diss}, respectively.
Finally, our main findings are summarised in $\S$\ref{sec:con}.

% We provide a new method to study both contributions, and reveal that inner dynamics are strongly We-dependent, while the outter depend only on the characteristics of the turbulent flow. The independence of the outter dynamics on
% the Weber number pave the way to the development of physically meaningful model of breakup. We stress that, as the We decreases, the contribution of the outter 
% dynamics becomes less relevant with respect to the inner dynamics, suggesting that the time-scale of the breakup process increases.

% \section{Methods}

\section{The exchange between kinetic energy and surface energy}
\label{sec:11}

We consider the Navier--Stokes (NS) equations coupled to the Cahn--Hilliard (CH) equation,
\begin{equation}
\begin{aligned}
\rho(\partial_t u_i + u_j \partial_j u_i) &= -\partial_i \tilde p + 2\partial_{j}\mu S_{ij} + f_i  - c\partial_i \phi, \\ 
\partial_t c + u_j\partial_j c &= \kappa\partial_{kk}\phi,
\end{aligned}
\label{eq:nse}
\end{equation}
which, together with the incompressibility constraint, $\partial_i u_i=0$, describe the evolution of an immiscible binary mixture of incompressible fluids \citep{jacqmin1999calculation}.
Here $u_i$ is the $i$-th component of the velocity vector, $\tilde p$ is a modified pressure, $S_{ij}=\tfrac{1}{2}(\partial_i u_j + \partial_j u_i)$ is the rate-of-strain tensor and $f_i$ is a body-force term per unit volume acting on the large scales to sustain turbulence.
Repeated indices imply summation, and we consider periodic boundary conditions.
The concentration of each component in the mixture is represented by $c$, 
where $c=\pm1$ are the pure components. The density, $\rho$, and the dynamic viscosity, $\mu$, of the fluid mixture depend on $c$, and the immiscibility is modelled through a chemical potential, 
\begin{equation}
\phi = \beta(c^2 - 1)c -  \alpha\partial_{kk}c.
\end{equation}
The true pressure is related to the modified pressure by $p=\tilde p + c\phi -\beta/4(c^2 - 1)^2 + \alpha/2(\partial_i c)^2$ \citep{jacqmin1999calculation}.
The action of interfacial forces in the momentum equation is represented by $c\partial_i\phi$, which 
is derived from
physical energy-conservation arguments,
and consistently reproduces the linear relation between surface tension forces and the local curvature of the interface \citep{jacqmin1999calculation}.
The numerical parameters $\alpha$ and $\beta$ determine the typical width of the fluid-fluid interface, 
\begin{equation}
\delta=4\sqrt{2\alpha/\beta},
\end{equation}
and the mobility, $\kappa$, determines its typical relaxation time. When these parameters are fixed appropriately, $\kappa\sim\delta^2$  \citep{magaletti2013sharp}, 
the interface is consistently close to the equilibrium profile, $c_{eq}(\xi)=\tanh \left(4 \xi/\delta\right)$, where $\xi$ is the spatial coordinate in the direction normal to the surface tangent plane. The surface tension reads
\begin{equation}
\sigma=\alpha\int_{-\infty}^{+\infty}\big(\partial_i c_{eq}\big)^2\mathrm{d}\xi=\frac{4}{3\sqrt{2}}\sqrt{{\alpha\beta}}.
\label{leq:surf}
\end{equation}
\com{The limits of the integral indicate that it is taken between large distances (formally infinite) compared to the interface thickness, $\delta$.
In practice, $95\%$ of the integral in (\ref{leq:surf}) is contained in $\xi \in[-\delta/2,\delta/2]$}.

\subsection{The evolution equations of the kinetic energy and the free energy}

We obtain the evolution equation of the kinetic energy of the flow by taking the dot product of the NS equations with $u_i$. 
%\ma{Marc: I'm not familiar with the expression contracting, I normally use "by taking the dot product with", but if contracting is common, just keep it.}
For convenience, we use the transformation $-c\partial_i \phi=\phi\partial_i c - \partial_i(c\phi)$, and add the second term in the
right-hand side to a new modified pressure, $p'=\tilde p - c\phi$. 
Then the equation reads
\begin{equation}
\partial_t e + u_j\partial_j e= \partial_i ( -u_i p' + 2\mu u_j S_{ij}) - 2\mu S_{ij}S_{ij} + u_i\phi\partial_i c + u_i f_i,\\ 
\label{eq:energy1}
\end{equation}
where $e=1/2 \rho u_iu_i$ is the turbulent kinetic energy per unit mass.
% \com{The density does not change along Lagrangian trajectories, 
% and $\rho(\partial_t u_iu_i + u_j\partial_j u_iu_i=\partial_t e + u_j\partial_j e$.}
% Considering periodic boundary conditions, or in the absence of boundary fluxes, 
The only terms contributing on average to the total kinetic energy budget are the local kinetic energy dissipation, $2\mu S_{ij}S_{ij}$, the power input, $u_i f_i$, %
and the energetic exchange between the kinetic energy and the surface energy, $u_i\phi\partial_i c$.
%This term acts %is only effective 
%in the fluid-fluid interface and is also present, with changed sign, in the evolution equation of the free energy, which is obtained 

By multiplying the CH equation by the chemical potential, we obtain
\begin{equation}
\partial_t h + u_j\phi\partial_j c = \kappa\phi\partial_{kk}\phi,
\label{eq:free_energy}
\end{equation}
where
\begin{equation}
h=\beta/4(c^2 - 1)^2 + \alpha/2 (\partial_k c)^2,
\end{equation}
is the free energy per unit volume.
\com{When integrated across the interface thickness,
this quantity transforms into the energy per unit area of the interface,}
\begin{equation}
\sigma=\int_{-\infty}^{+\infty} h(\xi) \mathrm{d}\xi,
\label{surface}
\end{equation}
where we have assumed an equilibrium profile of the phase field, $c_{\text{eq}}$, and the integral is performed as in (\ref{leq:surf}).

%%
%%
%We seek to reveal the fundamental mechanism that drives %operates 
%the energetic exchange between the kinetic energy and the surface energy

We now transform (\ref{eq:free_energy}) into an advection equation for $h$ by decomposing the product $\phi\partial_i c$ and operating on the partial derivatives.
We find the relation 
%\ma{Marc: maybe we should indicate what kind of operation is applied}
%%
\begin{equation}
\begin{aligned}
% -c\partial_i\phi  &= - \partial_i ( \phi c )  +  \phi \partial_i c \\
% \phi \partial_i c &=  (\beta(c^2 -1)c - \alpha \partial_{kk}c)\partial_i c \\ 
% \phi \partial_i c &=  \partial_i (\beta/4(c^2 - 1)^2) - \alpha \partial_{kk} c \partial_i c \\
% \phi \partial_i c &=  \partial_i (\beta/4(c^2 - 1)^2 + \alpha/2 (\partial_k c)^2) -\alpha\partial_k(\partial_k c \partial_i c). \\
% % \partial_t u_i&= - \alpha\partial_k \tau_{ik} - \partial_i p'
\phi \partial_i c &=  \partial_i h -\alpha\partial_j\tau_{ij},
\end{aligned}
\end{equation}
where $\tau_{ij}=\partial_i c\partial_j c$ is a Korteweg stress tensor.
Substituting this expression into the kinetic energy equation and the free energy equation, we obtain
%%
% \begin{equation}
\begin{align}
\partial_t e + u_j\partial_j e &= \partial_i \Psi_i - 2\mu S_{ij}S_{ij} - \alpha u_i\partial_j \tau_{ij} + u_i f_i,\label{eq:1}\\
\partial_t h + u_j\partial_j h &= \kappa\phi\partial_{kk}\phi + \alpha u_i\partial_j \tau_{ij},\label{eq:2}
\end{align}
% \end{equation}
%%
where $\Psi_i=u_i (-p' + h) + 2\mu u_j S_{ij}$, and $-p'+ h = -p + \alpha(\partial_i c)^2$.
The free energy equation has been transformed into an advection equation, where the first term in the right-hand side represents the diffusive and dissipative action of the chemical potential,
and the second term the interaction of the fluid-fluid interface with the velocity field through the stress tensor, $\tau_{ij}$.
This term also appears in the kinetic energy equation, and represents the conservative energetic
exchange between the free energy of the interface and the kinetic energy of the flow.

By taking the volume average of (\ref{eq:1}) and (\ref{eq:2}), we obtain an equation for the evolution of the total kinetic energy and the surface energy, $\mathcal E=\langle e \rangle_V$ and $\mathcal H=\langle h \rangle_V$,
%\begin{equation}
%%
\begin{align}
d_t \mathcal{H}&= \kappa\langle\phi\partial_{kk}\phi\rangle_V - \langle \alpha u_i\partial_j \tau_{ij} \rangle_V, \label{eq:ener31}\\
d_t \mathcal{E}&= 2\langle \mu S_{ij}S_{ij}\rangle_V + \langle \alpha u_i\partial_j \tau_{ij} \rangle_V + \langle u_i f_i\rangle_V, \label{eq:ener32}
\end{align}
\com{where the spatial fluxes vanish after averaging provided that there are no fluxes through the boundaries. This is the case for the periodic boundary conditions.} 
Note that by virtue of (\ref{surface}), the total surface energy is $\mathcal H=\sigma A$, where $A$ is the surface of the fluid-fluid interface.
On average, $\alpha u_i\partial_j \tau_{ij}$ is the only term responsible for the exchange between the surface energy and kinetic energy.

% Note that these equations are valid independently of the physical properties of the fluids in the mixture.
% In the case of different viscosities, the balance of tangential stresses produces a discontinuity in the rate-of-strain tensor across the interface (in the sharp-interface limit), but this discontinuity only affects the off-diagonal components of the rate-of-strain tensor in a frame fixed to the interface normal vector, which do not enter $\vartheta$ \citep{dopazo2000vorticity}. Thus the expression for $\vartheta$ remains valid and has equal value at both sides of the interface. 

%Let us note that $d_t \mathcal H=\kappa\langle\phi\partial_{kk}\phi\rangle_V + \langle\alpha u_i\partial_j \tau_{ij}\rangle_V$, where $\mathcal H=\langle h \rangle_V$ is the total surface energy, and $\langle\cdot\rangle_V$ denotes volume average. Since $\langle\phi\partial_{kk}\phi\rangle_V<0$, $\alpha u_i\partial_j \tau_{ij}$ is the only term responsible for the increments of the surface energy.

%\comav{Include equation 2.11 here}
%\comma{To me it makes more sense, for the sake of presentation, not to divide by the density in the kinetic energy equation. I prefer both equations in units of energy per volume. This way it is also clear that what is subtracted in one equation is added to the other.\\
%We could also consider to volume-average the equations to make the arguments clearer, i.e. work with total kinetic and free energy.}

\subsection{The physical mechanism of the surface energy variations}

To gain a better physical understanding of the energy-exchange term, 
we further expand it into
\begin{equation}
\alpha u_i\partial_j \tau_{ij}=\alpha \partial_j \left(u_i\tau_{ij}\right) - \alpha S_{ij}\tau_{ij}.\\ 
\label{eq:10}
\end{equation}
The first term in the right-hand side represents the divergence of a flux and vanishes in the mean, so that only the 
second term in the right-hand side has a net non-zero contribution.
Furthermore, due to the symmetric form of $\tau_{ij}$, only the symmetric part of the velocity gradient tensor, the rate-of-strain tensor $S_{ij}$, interacts with $\tau_{ij}$. Considering that the components of the vector normal to the surface tangent plane are $n_i=\partial_i c/\gamma$, 
where $\gamma=\sqrt{(\partial_k c)^2}$, we rewrite the Korteweg stress tensor as $\tau_{ij}=-\alpha\gamma^2n_in_j$, and the exchange term as $-\alpha \gamma^2 n_i S_{ij} n_j$.
This term describes the change in the free energy per unit volume. We transform it into a change of energy per unit area
by integrating it normal to the interface, which yields 
% By integrating normal to the interface we obtain the equation for the surface energy, $h_s$,
%%
% \begin{equation}
% \partial_t h_s + u_j\partial_j h_s=D_s + \vartheta,
% \end{equation}
%%
% where
\begin{equation}
\vartheta= - \sigma n_i S_{ij} n_j.
\label{eq:stretching}
\end{equation}
Here we have assumed that the interface is in equilibrium, so that (\ref{leq:surf}) holds, 
and that neither $\boldsymbol n$ nor $S_{ij}$ change substantially across the interface width. 
% \com{which is where $(\partial_i c)^2$ is non-zero.}
Both assumptions are fulfilled in the sharp-interface limit \citep{magaletti2013sharp}. 

Introducing the above analysis in the evolution equations for the kinetic energy and the surface energy, 
we obtain 
%\begin{equation}
%%
\begin{align}
d_t \mathcal{H}&= \kappa\langle\phi\partial_{kk}\phi\rangle_V + \langle \vartheta \rangle_S, \label{eq:ener311}\\
d_t \mathcal{E}&= 2\langle \mu S_{ij}S_{ij}\rangle_V - \langle \vartheta \rangle_S + \langle u_i f_i\rangle_V, \label{eq:ener322}
\end{align}
where $\langle\cdot\rangle_S$ denotes the integral over the fluid-fluid surface.
On average, $\vartheta$ is the only term responsible for the exchange of surface and kinetic energies.
These equations are valid independently of the physical properties of the fluids in the mixture.
In the case of different viscosities, the balance of tangential stresses produces a discontinuity in the rate-of-strain tensor across the interface (in the sharp-interface limit), but this discontinuity only affects the off-diagonal components of the rate-of-strain tensor in a frame fixed to the interface normal vector, which do not enter $\vartheta$ \citep{dopazo2000vorticity}. Thus the expression for $\vartheta$ remains valid and has equal value at both sides of the interface.

%This tensor has only one non-zero eigenvalue $\ell_1=d^2$,
%whose eigenvector points normal to the surface, and two zero eigenvalues, with eigenvectors parallel to the tangent plane of the surface. 
%Let us consider a frame of reference fixed to the interface, where
%Projecting the $S_{ij}$ tensor on the set of eigenvectors of $\tau_{ij}$ and considering the negative sign of $\alpha S_{ij}\tau_{ij}$, we have 

The expression in (\ref{eq:stretching}) resembles the vortex stretching term in the evolution equation of the enstrophy (the square of the vorticity vector),
which is responsible for its amplification \citep{betchov1956inequality}. 
By analogy, $\vartheta$ describes the stretching or contraction of the interface width by the rate-of-strain tensor.
%The negative sign indicates that the fundamental mechanism that generates surface energy is the compression \ma{Marc: do you mean extension?; see below}
%\comav{By compression and stretching I mean compression of the interface thickness which is equal to stretching of the surface area.}
%of the interface in the direction normal to the surface.
%
This result may be difficult to interpret from a physical and geometrical perspective,
because the interface between inmiscible fluids has a width of molecular scale, 
and its relaxation time is much faster than the time-scale of the velocity gradients.
In what follows, we give this term a physical interpretation.

\begin{figure}
\centering
\begin{tabular}{ll}
\hspace{0\textwidth}(a) \hspace{0.45\textwidth} (b) &\\
\includegraphics[width=1\textwidth]{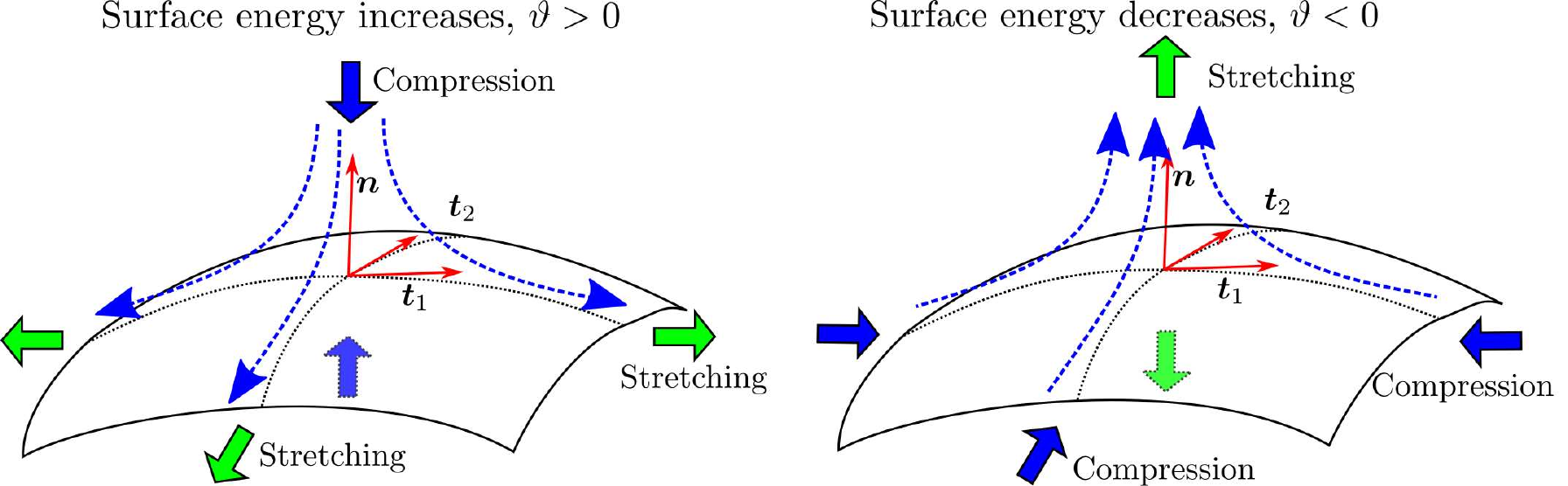}&\\
% \hspace{0\textwidth}(c)&\\
% \includegraphics[width=1\textwidth]{breakup.pdf}&\\
\end{tabular}
\caption{(a,b) Schematic representation of the mechanism that generates (a) positive and (b) negative increments of the surface energy due to the compression or stretching of the interface between two inmiscible liquids, where $\boldsymbol n$ is the orthogonal vector normal to the interface, $\boldsymbol t^1$ and $\boldsymbol t^2$ are vectors parallel to the surface, and $\vartheta=-\sigma n_i S_{ij} n_j$.
Blue and green arrows indicate the compressive and stretching directions of the rate-of-strain tensor at the surface, and dotted lines are the streamlines of the velocity field with respect to the surface. 
}
\label{fig:mech}
\end{figure}

For simplicity and consistency with the numerical simulations shown later in this paper, we have considered incompressible fluids.
Nevertheless, (\ref{eq:stretching}) holds regardless of the divergence of the velocity field.
Compressibility modifies both (\ref{eq:ener31}) and (\ref{eq:ener32}), since new fluxes and variables appear,
but it does not modify the structure of the exchange term, which remains the only net source of energy exchange.
To derive (\ref{eq:stretching}) from $\alpha u_i\partial_j \tau_{ij}$,
we have only assumed that the gradients of the velocity field take place at a scale much larger than the interface thickness. 
Thus, (\ref{eq:stretching}) is also valid for compressible flows as long as the velocity field does not contain any discontinuity (shock) in the interface. 
For the sake of generality we consider, in the following analysis, a compressible flow, and decompose the rate-of-strain tensor into a deviatoric and a volumetric part, $S_{ij}=S^d_{ij} + (P/3)I_{ij}$, where $I_{ij}$ is the identity tensor, and $P=S_{lk}I_{lk}$ is the local volumetric rate of expansion or contraction.
Since $S^d_{ij}I_{ij}=0$ and $I_{ij}n_jn_i=1$, we arrive at
\begin{equation}
S_{ij} n_j n_i=S^d_{ij} (n_j n_i - I_{ij}) + P/3.
\end{equation}
Now we reformulate (\ref{eq:stretching}) in terms of $P$ and of any pair of orthonormal vectors parallel to the surface, $\boldsymbol{t}^1$ and $\boldsymbol{t}^2$,
%%%
\begin{equation}
\vartheta= \sigma( t^1_k S^d_{kj} t^1_j + t^2_k S^d_{kj} t^2_j) - \sigma P/3, 
\end{equation}
%%%
where we have considered that, $n_j n_i - I_{ij}=-t^1_it^1_j - t^2_it^2_j$. The first term represents the growth rate of an infinitesimal surface area, $\delta A$, due to the deviatoric part of the rate-of-strain tensor, which reads 
\begin{equation}
\frac{1}{\delta A}\mathrm{d}_t\delta A= t^1_k S^d_{kj} t^1_j + t^2_k S^d_{kj} t^2_j.
\end{equation}
The second term is related to compressibility effects, and is proportional to the growth rate of an infinitesimal volume at the surface. 

In the case of an incompressible flow, in which $P=0$ and $S_{ij}=S^d_{ij}$,
\begin{equation}
\vartheta=\sigma( t^1_k S_{kj} t^1_j + t^2_k S_{kj} t^2_j),
\end{equation}
which shows that the surface energy increases as a consequence of the stretching of the surface area by the rate-of-strain tensor.
%This expression complies well with the intuitive notion that the surface must expand for the surface energy to increase. In fact, the increments of an infinitesimal area on the surface of the drop is .
The opposite mechanism is also possible, and
the decrements of the surface energy are related to the contraction of the surface area.
In figure \ref{fig:mech}(a,b) we show a schematic representation of these mechanisms for an incompressible flow. 

\section{Single-drop experiments in homogeneous isotropic turbulence}
\label{sub:nm}

The analysis presented in the previous section is valid for any configuration of the fluid-fluid interface, or flow regime. Hereinafter we apply our analysis to the dynamics of a single drop embedded in a homogeneous and isotropic turbulent flow.  

\subsection{Numerical method}

We consider two fluids with equal density and kinematic viscosity, 
and integrate (\ref{eq:nse}) in a triply periodic cubic domain of volume $L^3=(2\pi)^3$ by projecting the equations on a basis of $N/2$ Fourier modes in each direction, where $N=256$.
%The 
Non-linear terms %of the NS and CH equations 
are computed through a dealiased pseudo-spectral procedure, 
and a third-order semi-implicit Runge-Kutta scheme is used for the time integration, 
with a decomposition of the linear terms proposed by \citet{badalassi2003computation}. 
To sustain turbulence in a statistically steady state, we implement a linear body-force, $\widehat f_i=C_f\widehat{u}_i$, that is only applied to wavenumbers $k<2$, where $\widehat{\cdot}$ denotes the Fourier transform and $k$ is the wavenumber magnitude. 
The forcing coefficient, $C_f$, is set so that, at each time, the total kinetic energy per unit time injected in the system is equal to a constant, which is chosen so as to fix the numerical resolution, $k_{max}\eta=4$, where $\eta=(\nu/\varepsilon^4)^{3/4}$ is the Kolmogorov length scale and $k_{max}=N/3$ is the maximum wavenumber magnitude after dealiasing.
%The time scales associated to the Kolmogorov scale is $t_\eta=(\eta^2/\varepsilon)^{1/3}$.

The Reynolds number of the flow is $Re_\lambda=\lambda u'/\nu=58$, where $\lambda=\sqrt{15(\nu/\varepsilon})u'$ 
is the Taylor microscale, $u'=\sqrt{2\mathcal{E}/3}$ is the root-mean-square of the velocity fluctuations, and $\mathcal{E}=1/2\langle u_iu_i\rangle$ is the ensemble-averaged kinetic energy. The thickness of the fluid-fluid interface is set by the Cahn number $\mathrm{Ch}=({{\alpha}/{\beta L^2}})^{1/2}=0.012$, which for $N=256$ is appropriate to resolve the interface with a spectral Fourier basis \citep{chen1998applications}. 
%\textcolor{red}{Marc: we should say that for the Fourier method this is fine, by citing the paper of Shen}\\
The mobility $\kappa$ defines the Peclet number, $\mathrm{Pe}={u' L^2}/{\kappa\sqrt{\alpha\beta}}=3\mathrm{Ch}^{-2}$, which is fixed to ensure that the dynamics of the interface are consistent in the sharp-interface limit \citep{magaletti2013sharp}.
The time-step is set to $\varDelta t=0.04\mathrm{Ch}$.
Simulations have been performed on GPUs using a modified version of the spectral code described in \cite{cardesa2017}.

The code has been validated against \citet{shao2018}, and the results of this validation analysis are presented in appendix I.
In addition, the consistency of the numerical parameters, such as the time-step and the spatial resolution, have been checked. 
%\textcolor{red}{Marc: this sentence may not be enough. We should say we have not only validated the code, but also the resolution, time-step, etc.}

\begin{figure}
\centering
\begin{tabular}{ll}
% \hspace{0\textwidth}(a) \hspace{0.45\textwidth} (b) &\\
% \includegraphics[width=1\textwidth]{mechanism_2.pdf}&\\
\hspace{0\textwidth}&\\
\includegraphics[width=1\textwidth]{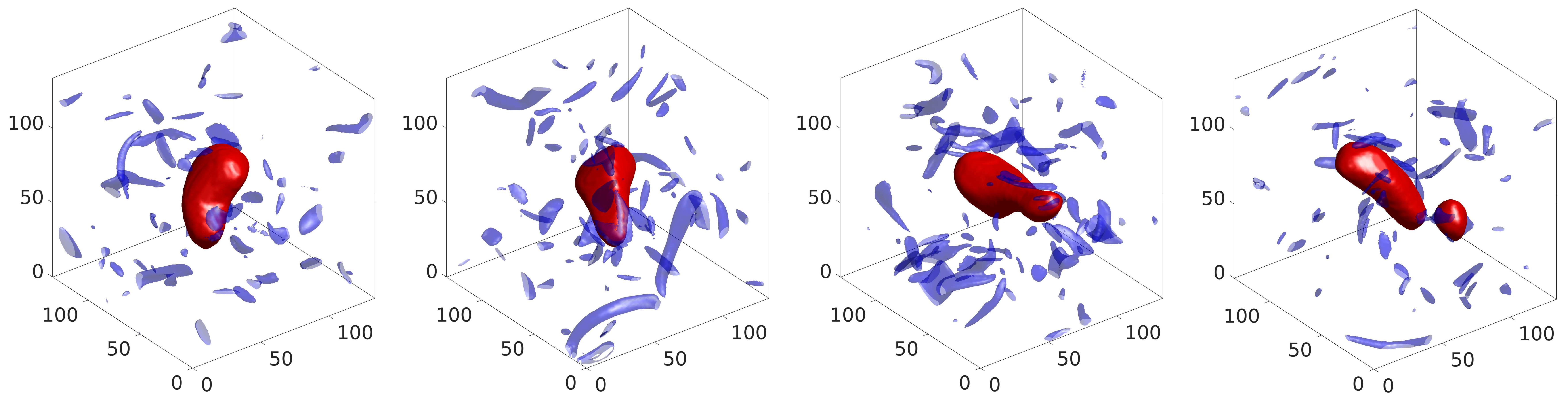}&\\
\end{tabular}
\caption{ Temporal evolution of a drop at $\mathrm{\it We}=1.8$.
The frame is fixed at the centre of the drop, and the time of the snapshots corresponds, from left to right, to $t/t_d=2.0$, $2.9$, $4.8$, and $5.1$. Blue isosurfaces denote vorticity with magnitude $|\boldsymbol \omega|=2.6\langle|\boldsymbol \omega|\rangle$, where the brackets denote the ensemble average.
The size of the computational box is marked in Kolmogorov units.
}
\label{fig:mech_drop}
\end{figure}

\begin{figure}
\centering
\begin{tabular}{lll}
\hspace{-10pt}(a)&
\hspace{-10pt}(b)&
\hspace{-15pt}(c)\\
\hspace{-10pt}\includegraphics[width=0.35\textwidth]{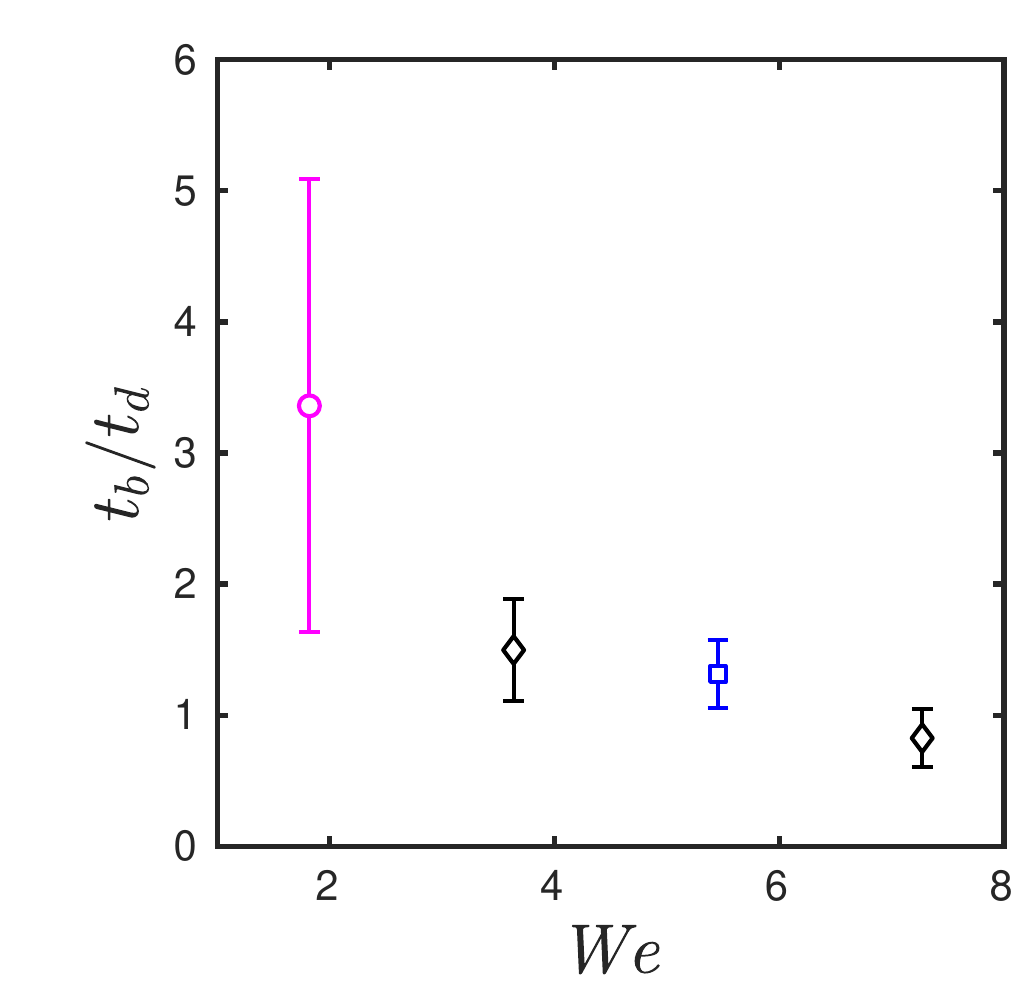}&
\hspace{-10pt}\includegraphics[width=0.35\textwidth]{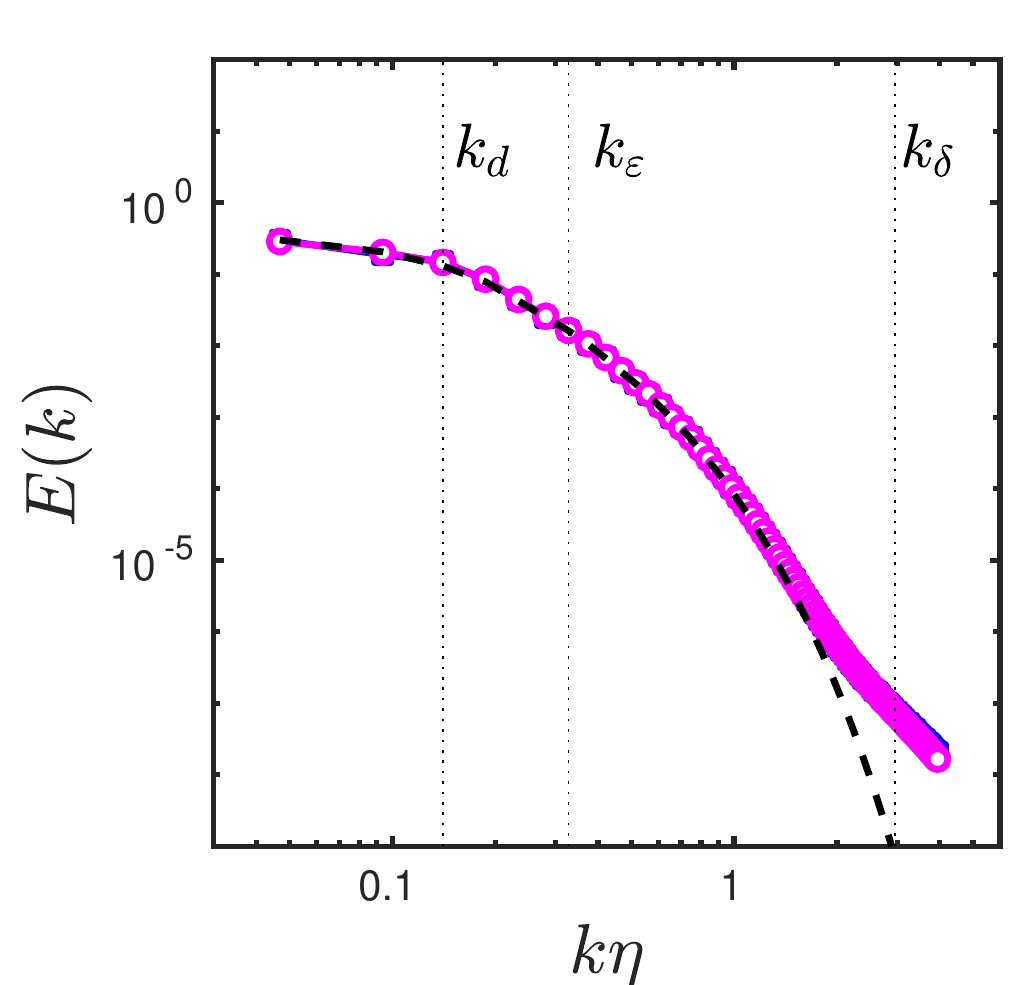}&
\hspace{-15pt}\includegraphics[width=0.35\textwidth]{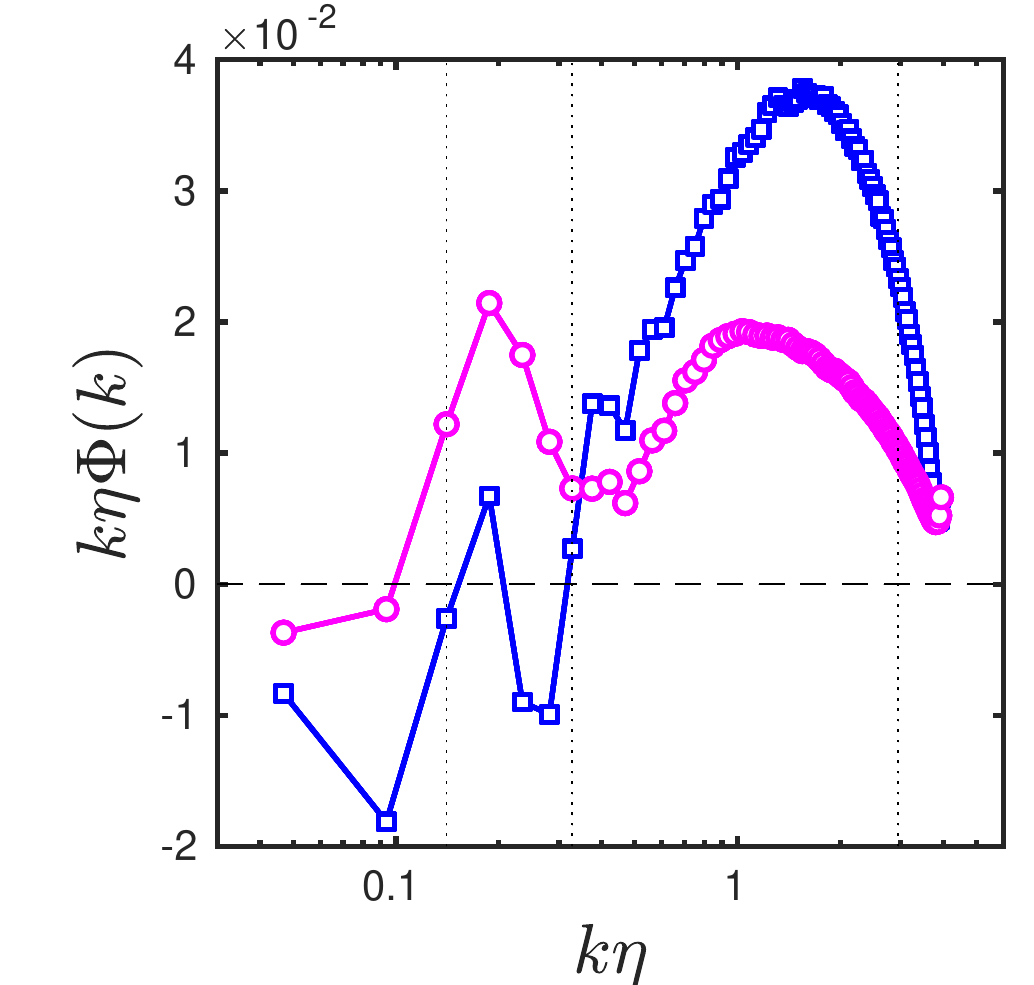}
\end{tabular}
    \caption{(a) Median time to breakup as a function of the $\We$. The upper and lower bars mark plus-minus the standard deviation. 
         \com{The style of the markers is only used for ease of reference for the data shown in the following panels.} 
         (b) Average kinetic energy spectra at $Re_\lambda=58$ for: 
         \protect\dashed, turbulent flow without droplet;    
         \protect\linescircle{magenta}, $\mathrm{\it We}=1.8$;
         \protect\linesquare{blue}, $\mathrm{\it We}=5.4$.
         Vertical dotted lines mark the scale (wavenumber) of the drop, $k_d=2\pi/d$, the scale where the spectral density of the kinetic energy dissipation is the highest, $k_\varepsilon$, and the scale of the interface, $k_\delta=2\pi/\delta$. 
         The spectra are averaged across the flow fields in the database.
         (c) Pre-multiplied production spectra, $k\eta\Phi(k)$. Lines as in (b). The total contribution of each wavenumber to the variation of the surface free energy is represented as the area below the curve. $\Phi$ is normalised with its average in each case.}
         %I have to find a proper way to normalise figure B.
         %} %
\label{spectra}
\end{figure}

\subsection{Initial conditions and drop size}

%We study the evolution and breakup of a single drop placed in isotropic turbulence.
We introduce a drop of diameter 
$d=\tfrac{1}{3}L=45\eta$ in a fully developed turbulent flow at time $t=0$, and integrate the governing equations \eqref{eq:nse} until the drop breaks at time $t_b$.
In figure \ref{fig:mech_drop},
we show a visualisation of a typical single-drop simulation. 
Since $d\gg\eta$, breakup is dominated by inertial forces and characterised by the Weber number, $\mathrm{\it We} = {\rho \varepsilon^{2/3} d^{5/3}}/{\sigma}$, and by a characteristic inertial time scale 
\begin{equation}
t_d=(d^2\varepsilon)^{1/3}.
\end{equation}
To statistically characterise drop deformation, we perform many independent single-drop simulations %experiments 
initialised with statistically independent turbulent flow fields.
Mass leakage \citep{Yue2007} leads to a slow but progressive reduction of the drop diameter, and to a time-dependent Weber number, $\We^\dagger(t)$, which decreases slightly through the simulations. However, this variation is small, and therefore we do not employ any special numerical approach to prevent it \citep{zhang2017flux,soligo2019mass}. 
We consider the effective Weber number of our simulations as the average Weber number at the average time of breakup, $\langle t_b\rangle$, i.e $\We=\langle \We^\dagger(\langle t_b\rangle)\rangle$. Here the brackets denote the ensemble average.
The difference between the time-dependent Weber number at the time of breakup, $\We^\dagger(t_b)$, and the effective Weber number, $\We$ is at most $\sim3\%$ in the worst cases.

We have performed simulations %experiments 
at four different effective $\We\in[1.8,7.4]$. 
For each $\We$, the number of simulations is approximately $100$,
which yields a total simulation time of between $300t_d$ and $1500t_d$.

From each simulation, we stored a sufficient number of full flow fields to have fully converged statistics. 
\com{We analysed flow fields at times at least $0.5t_d$ after the initial condition and before breakup.
We assume that in this time interval the evolution of the drop is statistically stationary.} 
\com{In this range of $\We$, breakup occurs in times of the order of the inertial time-scale of the drop, $t_d$.}
In figure \ref{spectra}(a), we show the median and the standard deviation of the time to breakup, $t_b$, as a function of the $\We$. As expected, lower values of the $\We$ yield  longer times to breakup. 

In what follows we show that although $d$ is comparable to the integral scale of the flow, the drop does neither %not 
modify the structure of the surrounding turbulence %or
nor resonates with the numerical box. The drop interacts mostly with scales smaller than $d$, indicating that the linear forcing used to sustain turbulence does not affect breakup. In figure \ref{spectra}(b), we show the average kinetic energy spectrum, $E(k)=2\pi k^2 \langle \widehat u \widehat u^* \rangle_k$, with and without an immersed drop. Here $\langle \cdot \rangle_k$ denotes averaging over modes 
with wavenumber magnitude $k$, and $\cdot^*$ represents the complex conjugate.
The energy spectra are similar for the flow with and without droplet above $k\eta\sim1$, suggesting a similar turbulent structure in those scales. To further corroborate this, we calculate the skewness and flatness factor of the longitudinal velocity derivatives, 
\begin{equation}
    \mathcal{F}_n=\frac{\langle (\partial_i u_i)^n \rangle}{\langle (\partial_i u_i )^2\rangle^{n/2}},
\end{equation}
where no summation is intended for repeated indices. We find that, away from the drop surface, $\mathcal{F}_3\approx-0.52$ and $\mathcal{F}_4\approx5.2$ independently of the Weber number, which are similar to the values for a simulation without the drop and in the expected range for the $Re_\lambda$ considered \citep{jimenez1993structure}. 

The good collapse of the energy spectra 
in the small wavenumbers (large scales) also suggests the absence of any resonances between the drop and the box, which could lead %leading 
to spurious large-scale dynamics. 
To examine how %prove that 
the large-scale forcing affects the dynamics of breakup, we study the pre-multiplied production spectra, $k\eta\Phi(k)$, shown in figure \ref{spectra}(c), where
%the effect that turbulent fluctuations have on the growth of the surface energy in scale space,
\begin{equation}
\Phi(k)=-4\pi k^2 \Re\langle\widehat{ \big( u_j\partial_j c\big)}{\widehat \phi}^* \rangle_k,
\end{equation}
describes the contribution of each scale to the changes
of the surface energy due to the deformation of the interface by the velocity field.
Note that the integral of $\Phi(k)$ across wavenumbers is equal to $\langle \vartheta\rangle_S$. %the pre-multiplied production spectra, $k\Phi(k)$, at different $\mathrm{We}$ shows 
Variations of the surface energy are due mostly to turbulent fluctuations at scales below the drop diameter,
whereas the contribution of larger scales is comparatively small.
In fact, the contribution of the scales affected by the forcing (with $k<2$) to the production of surface energy is slightly negative on average, confirming %corroborating
that the forcing does not contribute to deformation and breakup.

%\begin{table}
%\caption{Parameters of the ensembles of simulations performed.\comav{What should we put here?}}
%          \comma{we'll have to discuss what exactly to list here.}}
%\begin{center}
%\begin{tabular}{c c c c c c c c c c}
% & \multicolumn{3}{c}{$Re_\lambda=31$} & \multicolumn{3}{c}{$Re_\lambda=58$} & \multicolumn{3}{c}{$Re_\lambda=96$}  \\
%   \cline{2-4} \cline{5-7}\\
%$We$ & $\mathcal{N}_{LR}$ & $\mathcal{N}_{HR}$ & $\langle t_b %\rangle/t_d$ & $\mathcal{N}_{LR}$ & $\mathcal{N}_{HR}$ & $\langle t_b \rangle/t_d$ & $\mathcal{N}_{LR}$ & $\mathcal{N}_{HR}$ & $\langle t_b \rangle/t_d$ \\
% \hline
% $1.1$ & $1208$ &$-$&$8.78$& $1475$ & $109$& $8.06$ & $-$ & $180$  & $7.8$  \\  
% $1.8$ & $1208$ &$-$&$5.62$& $2001$ & $201$& $5.22$ & $-$ & $245$  & $5.79$  \\  
% $3.6$ & $1208$ &$-$&$4.58$& $1834$ & $201$& $4.31$ & $-$ & $248$  & $4.12$ \\   
% $5.4$ &        & \\
%s $7.3$ &        & 
 %  \hline
% \end{tabular}
% \end{center}
%\label{table_1}
%\end{table}

\section{Analysis of the energetic exchange in isotropic turbulence}
\label{sec:anal}

\begin{figure}
(a)\hspace{0.35\textwidth}(b)\hspace{0.25\textwidth}(c)\hspace{1\textwidth}\par
\centering
\includegraphics[width=0.9\textwidth]{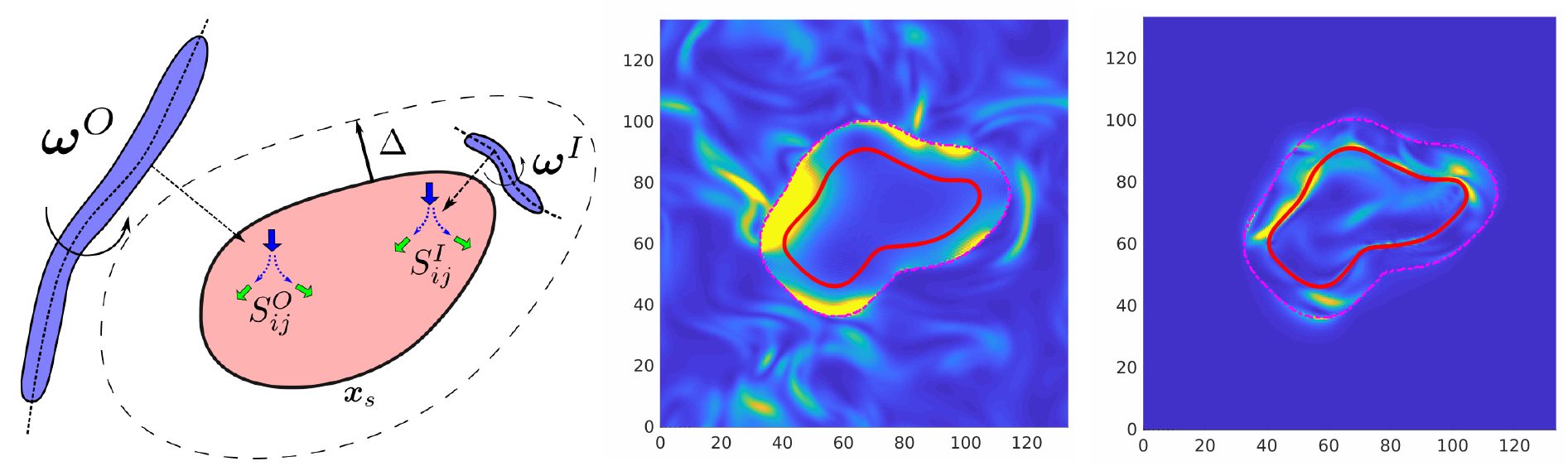}
\caption{(a) Decomposition of the
rate-of-strain tensor in contributions due to eddies a distance $\varDelta$ from the drop surface, denoted as $S^O_{ij}$,
and a distance $\varDelta$ close to the surface (including the inside of the drop), denoted as $S^I_{ij}$.
$\boldsymbol x_s$ marks the drop surface. 
(b,c) Magnitude of the (b) outer and (c) inner rate-of-strain tensor, $2S^O_{ij}S^O_{ij}$ and $2S^I_{ij}S^I_{ij}$, 
in a direct numerical simulation for $\varDelta=0.19d$ and $\We=5.4$.
The solid red line marks the drop surface, and the dashed magenta line a distance $\varDelta=0.19d$ from the surface. 
\com{In (b), the outer rate-of-strain tensor at a distance smaller than $\varDelta=0.19d$ from the drop surface has been multiplied by $10$ to ease visualisation.}
Axes in Kolmogorov units.}
\label{fig:mech00}
\end{figure}

\subsection{Local and non-local surface stretching}
\label{sec:decomp}

Despite the simplicity of the energetic exchange described in $\S$\ref{sec:11}, 
the coupling between the dynamics of the drop surface and the surrounding turbulence is bidirectional and highly non-linear. 
The rate-of-strain tensor generates surface contraction or expansion, 
and, at the same time, the surface dynamics generates straining motions. 
For instance, as a deformed drop relaxes toward a spherical shape,
the surface energy follows damped oscillations before settling into equilibrium (see appendix I).
Although these changes of the surface energy are produced by the stretching and compression of the surface, they are drop-induced and not necessarily related to drop breakup.
%which requires in general 
%that the surface energy increases due the action of the %ssurrounding turbulence. 

To separate the dynamics of the interface from the action of the surrounding turbulence,
we split the surface-stretching term into contributions related to eddies close to (including those inside) the drop, and far from the drop. 
Here we consider eddies as patches of swirling fluid, and associate them with the vorticity field. The kinematic relations that hold between the vorticity vector and the rate-of-strain tensor imply that eddies away from the drop stretch its surface by non-local effects \citep{ohkitani1995nonlocal,hamlington2008local}.
Let us consider a distance $\varDelta$, and separate the vorticity field in $\varDelta$-outer and $\varDelta$-inner eddies. 
The former are defined as eddies a distance $\varDelta$ or further from the drop surface, while the later correspond to eddies that are closer than $\varDelta$ or inside
the drop surface.
We define the vorticity field at a distance further than $\varDelta$ from the drop surface as
\begin{equation}
\boldsymbol \omega'^O=G(\boldsymbol x;\varDelta) \boldsymbol \omega, 
\end{equation}
where $\boldsymbol \omega=\nabla \times \boldsymbol u$ is the vorticity vector, and
\begin{equation}
% \begin{aligned}
G(\boldsymbol x;\varDelta)=
\begin{cases}
1 \text{ if $|\boldsymbol x - \boldsymbol x_s|_s>\varDelta$ and $\boldsymbol x \in \mathscr{O}$}, \\
0 \text{ if otherwise},
\end{cases}
% \end{aligned}
\end{equation}
is a kernel that truncates the vorticity field at a distance $\varDelta$ from the drop surface. Here $\boldsymbol x_s$ defines the surface of the drop, $|\boldsymbol x - \boldsymbol x_s|_s$ is the shortest Euclidean distance from $\boldsymbol x$ to the drop surface, and $\mathscr O$ comprises all the points outside the drop.
Let us note that $\boldsymbol \omega'^O$ does not define a vorticity field because it does not locally fulfil $\nabla\cdot\omega^O=0$, \com{where the truncation is performed}. We thus project it into the closest divergence-free field, $\boldsymbol \omega^O=\boldsymbol \omega'^O - \nabla \psi$, by solving
\begin{equation}
\nabla^2\psi=\nabla\cdot\boldsymbol\omega'^O,
\end{equation}
with periodic boundary conditions. We calculate the stretching induced 
on the drop by the eddies away from its surface from the Biot--Savart law \citep{ohkitani1995nonlocal}.
By taking the curl of the vorticity and considering that $\nabla\cdot\boldsymbol  u^O =0$, we obtain the following equation %We apply the rotor operator to the vorticity field and the invoke incompressibility of $\boldsymbol u^O$, leading to
\begin{equation}
\nabla^2\boldsymbol u^O=-\nabla\times\boldsymbol \omega^O,
\end{equation}
which, when solved with periodic boundary conditions, provides the rate-of-strain tensor produced by $\varDelta$-outer eddies, 
$S^O_{ij}=\tfrac{1}{2}(\partial_j u_i^O + \partial_i u_j^O)$, and by $\varDelta$-inner eddies, $S_{ij}^I=S_{ij} - S_{ij}^O$.

The stretching of the drop surface by $\varDelta$-outer and $\varDelta$-inner eddies are 
\begin{equation}
\begin{aligned}
\vartheta^O&=-\sigma n_iS^O_{ij}n_j,\\
\vartheta^I&=-\sigma n_iS^I_{ij}n_j,
\end{aligned}
\end{equation}
respectively, where the rate-of-strain tensor is evaluated at the surface of the drop.
%%
% \begin{equation}
% \vartheta^I=-n_iS^I_{ij}n_j
% \end{equation}
% %%
 %Here $\vartheta^O=-n_iS^O_{ij}n_j$, where the rate-of-strain tensor is evaluated at the points in the surface of the drop, 
%represents the stretching of the drop surface due to vorticity at distances larger than $\varDelta$ from the drop surface,
%
A schematic representation of this decomposition 
and its application to a snapshot of a direct numerical simulation are shown in figures \ref{fig:mech00}(a)--(c).

\subsection{Stretching by inner and outer eddies}
\label{sec:decomp2}

In this section, we obtain a value of $\varDelta$ for which the separation in $\varDelta$-inner and $\varDelta$-outer eddies is physically meaningful and practical. 
We aim to separate the surface stretching into a contribution due to eddies close to the drop surface, which are affected by surface dynamics or by the material properties of the drop, and another contribution due to eddies far from the surface, which are independent of surface dynamics or the material properties of the drop.

A practical approach is to find the smallest $\varDelta$ for which the $\varDelta$-outer eddies and the surface are not coupled, i.e, the $\varDelta$-outer stretching is independent of surface dynamics. In addition, this approach identifies the smallest region close to the drop surface where eddies are affected by surface dynamics.  
We denote the $\varDelta$-inner and $\varDelta$-outer eddies identified with this criterion simply as inner and outer eddies.

\begin{figure}
\centering
\begin{tabular}{ll}
(a)\vspace{-10pt}&
(b)\\
\includegraphics[width=0.5\textwidth]{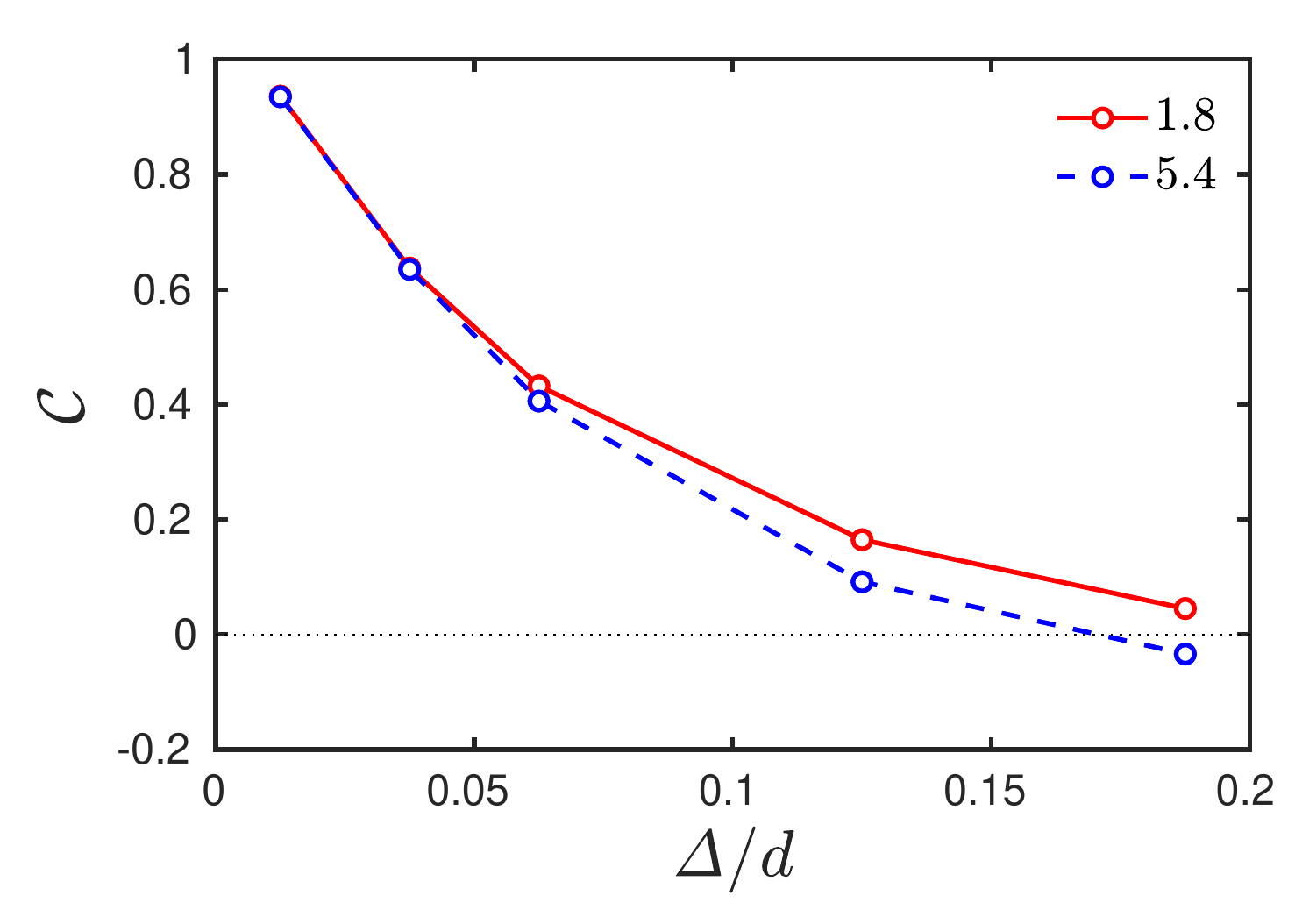}&
\includegraphics[width=0.5\textwidth]{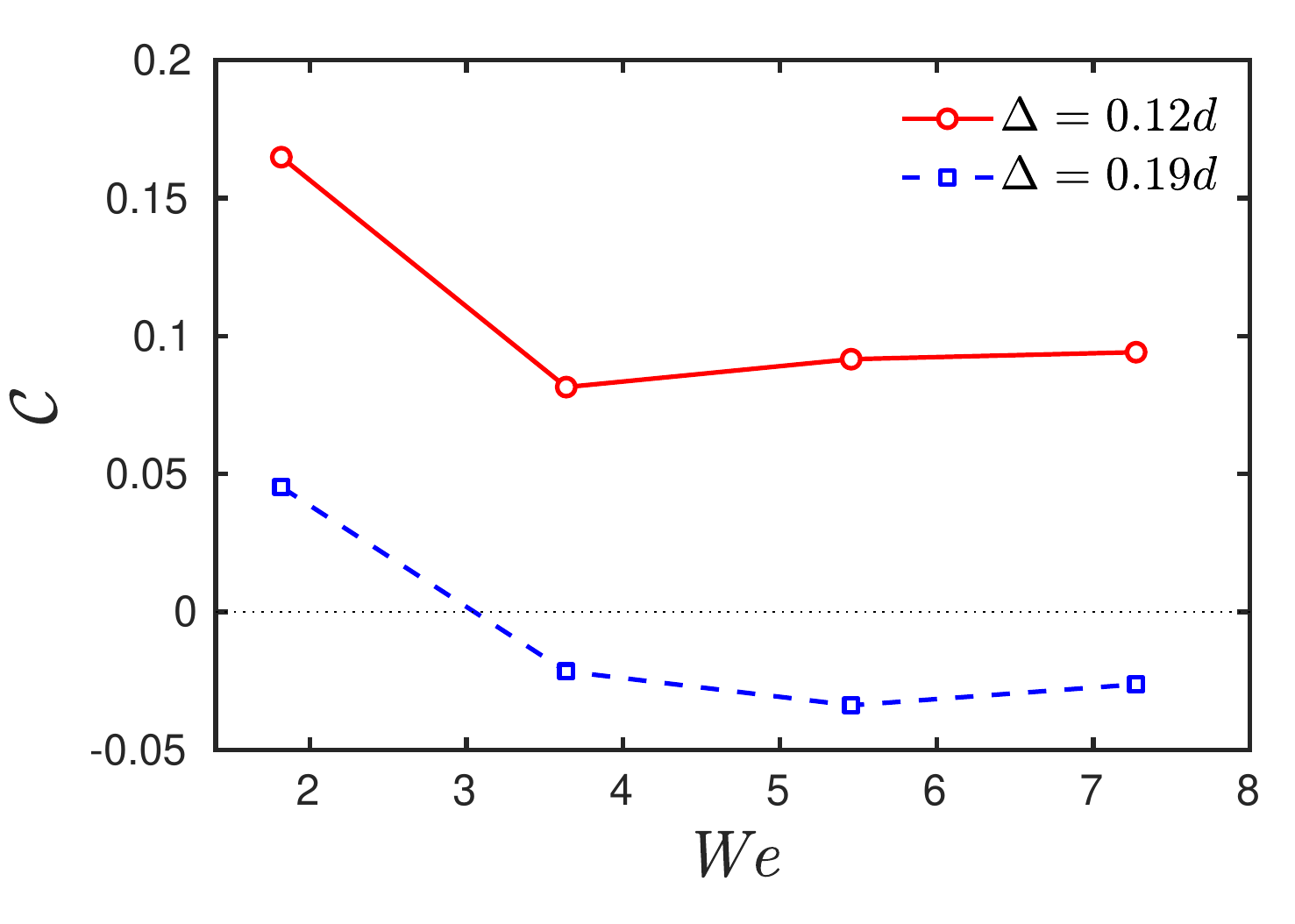}\\
\end{tabular}
\caption{
(a,b) Correlation coefficient, $\mathcal C$, between the stretching by inner eddies, $\vartheta^I$, and outer eddies, $\vartheta^O$, as a function of (a) $\varDelta$ and (b) $\We$.}
\label{fig:mech1}
\end{figure}

To find the $\varDelta$ that separates inner and outer eddies, we use the correlation coefficient between $\vartheta^O$ with $\vartheta^I$, 
\begin{equation}
    \mathcal C=\frac{\langle \overline{\vartheta^I}\cdot \overline{\vartheta^O}\rangle}{\sqrt{\langle {\overline{\vartheta^I}}^2\rangle\langle {\overline{\vartheta^O}}^2\rangle}},
\end{equation}
where the bar denotes quantities without their ensemble average, $\overline{\vartheta}=\vartheta-\langle \vartheta\rangle$.
We evaluate $\mathcal C$ for different $\We$ and $\varDelta$, 
and show the results in figure \ref{fig:mech1}(a,b). 
The correlation coefficient is close to unity for $\varDelta$ close to zero,
and decays fast with increasing $\varDelta$. For $\varDelta=0.12d$, $\mathcal C\approx0.2$, and for $0.19d$ it drops close to zero.
In figure \ref{fig:mech1}(b), we show that the correlation coefficient 
for $\Delta=0.12d$ and $\Delta=0.19d$ is close to zero for all $\We$, indicating that $0.12d<\varDelta<0.19d$ is a reasonable estimate of the distance that separates outer and inner eddies 
in the range of $\We$ considered here. This separation corresponds in Kolmogorov units to $6\eta<\varDelta<9\eta$.

In the following, we will study the contribution of inner and outer eddies to the surface stretching.
For simplicity, we name these contributions as the inner and outer stretching, respectively.

% which we denote for simplicity as
% inner and outer stretching.
% For simplicity, we name these contributions as the inner and outer stretching, respectively.

% We call them `inner' and `outer' eddies respectively. In this section, we show that in fact a separation in surface-driven, or `inner', and surface-independent, or `outer', eddies takes place at a small distances (compared to the drop diameter) from the drop surface. 

%In the following sections, we will further justify this decomposition by showing the very different statistics of inner and outer stretching. 

%$\varDelta\sim6\eta\sim0.12d$.
%In this section, we will provide an approximate limit to separate both classes of eddies, 
%and statistically characterise their contribution to the surface stretching. 

\subsection{Statistics of the surface stretching}

%In this section, we characterize the statistics of the inner and outer stretching, and their dependence with the $\We$.

\begin{figure}
\centering
\begin{tabular}{lll}
\hspace{-10pt}(a)&
\hspace{-15pt}(b)&
\hspace{-15pt}(c)\\
\hspace{-10pt}\includegraphics[width=0.35\textwidth]{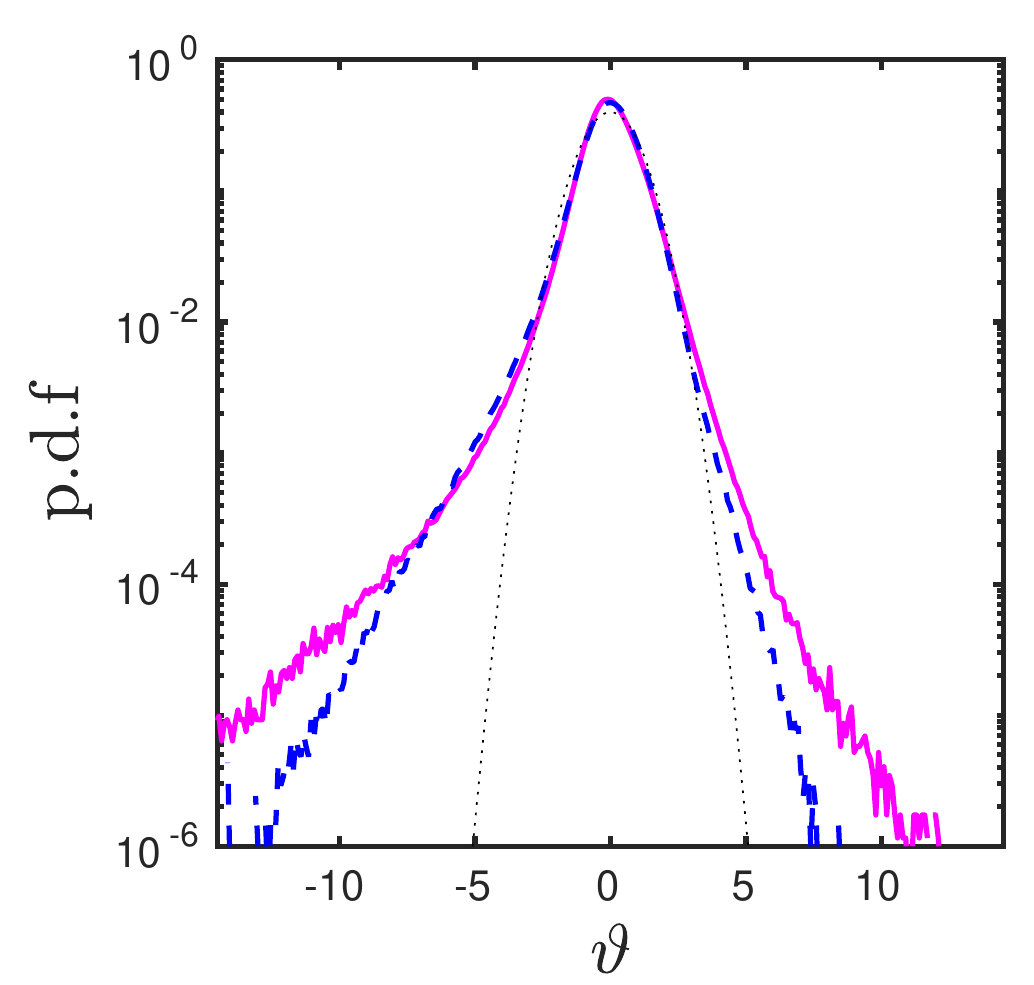}&
\hspace{-15pt}\includegraphics[width=0.35\textwidth]{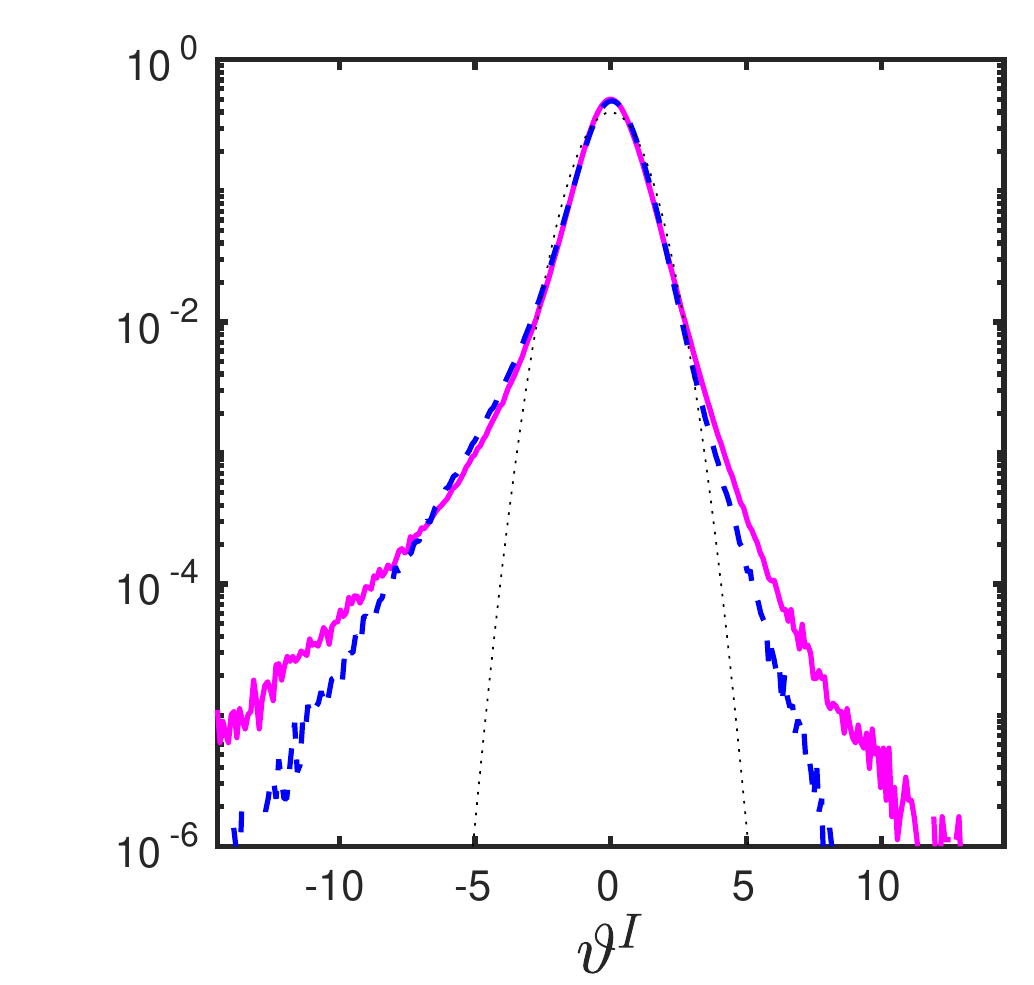}&
\hspace{-15pt}\includegraphics[width=0.35\textwidth]{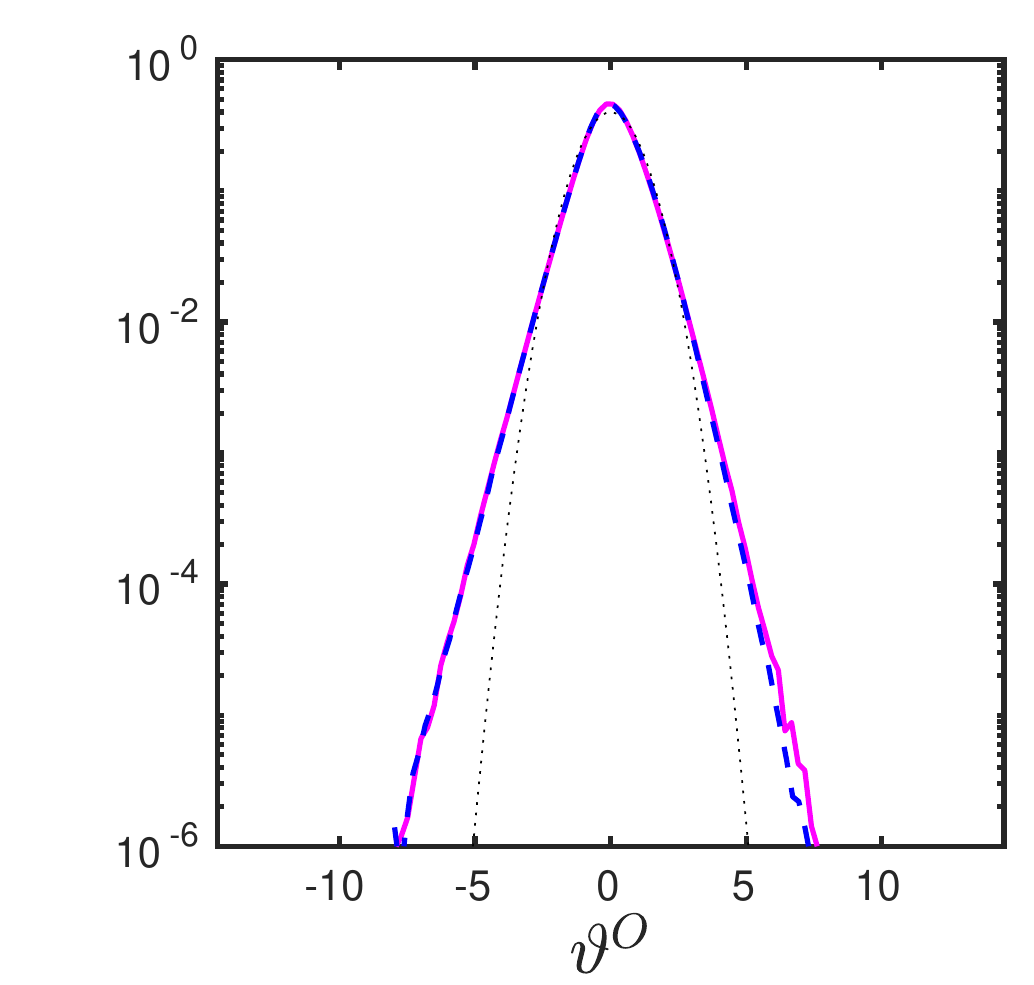}
\end{tabular}
\caption{Probability density function of the local surface stretching due to (a) the full flow, $\vartheta$, (b) the inner eddies, $\vartheta^I$,
and (c) the outer eddies, $\vartheta^O$, for $\varDelta=0.12d$ and at $\We=$: {\color{magenta}\protect\solid}, 1.8; {\color{blue}\protect\dashed} 5.4.
Quantities plotted without the mean and divided by their standard deviation.}
\label{fig:prob}
\end{figure}

% Here we analyse the statistics of the outer and inner stretching.
In figure \ref{fig:prob}(a)--(c), we show the probability density function (p.d.f) of $\vartheta$, $\vartheta^O$, and $\vartheta^I$,
measured at the surface of the drop for $\We=1.8$ and $5.4$, and $\varDelta=0.12d$. Results are similar for $\varDelta=0.19d$.
We find that $\vartheta$ and $\vartheta^I$ are very similar and display fairly fat-tailed distributions which depend on the $\We$. 
We suggests that these tails are a consequence of surface tension forces produced by the oscillations of the interface.
On the other hand, the p.d.f's of $\vartheta^O$ are more symmetric, closer to a Gaussian. 
Their good collapse at two different $\We$ indicates that the outer stretching is not affected by surface dynamics. 
These results corroborate that the value of $\varDelta$ used to decompose the flow in inner and outer eddies is adequate.

\begin{figure}
\centering
\begin{tabular}{ll}
(a)&
(b)\\
\includegraphics[width=0.5\textwidth]{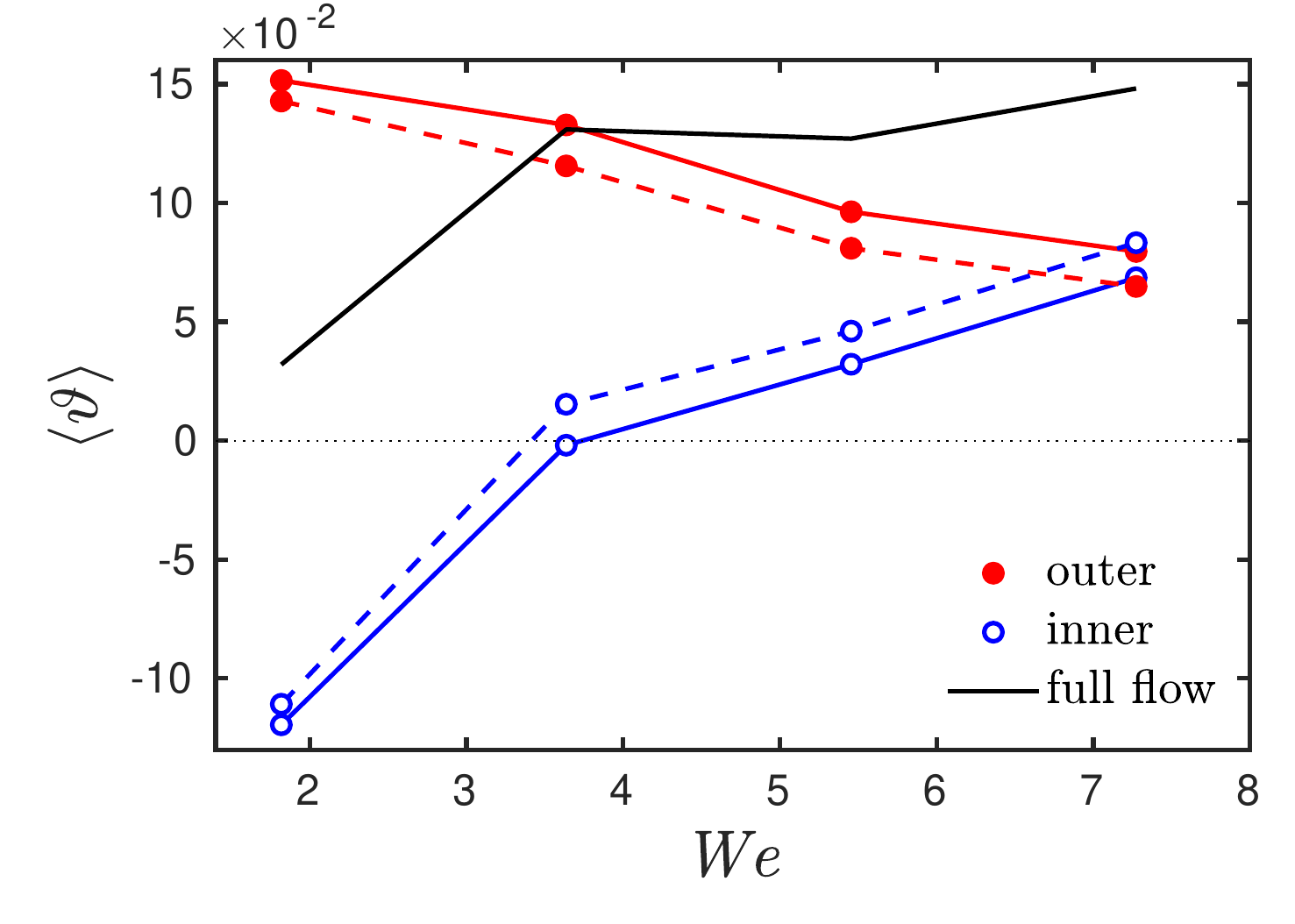}&
\includegraphics[width=0.5\textwidth]{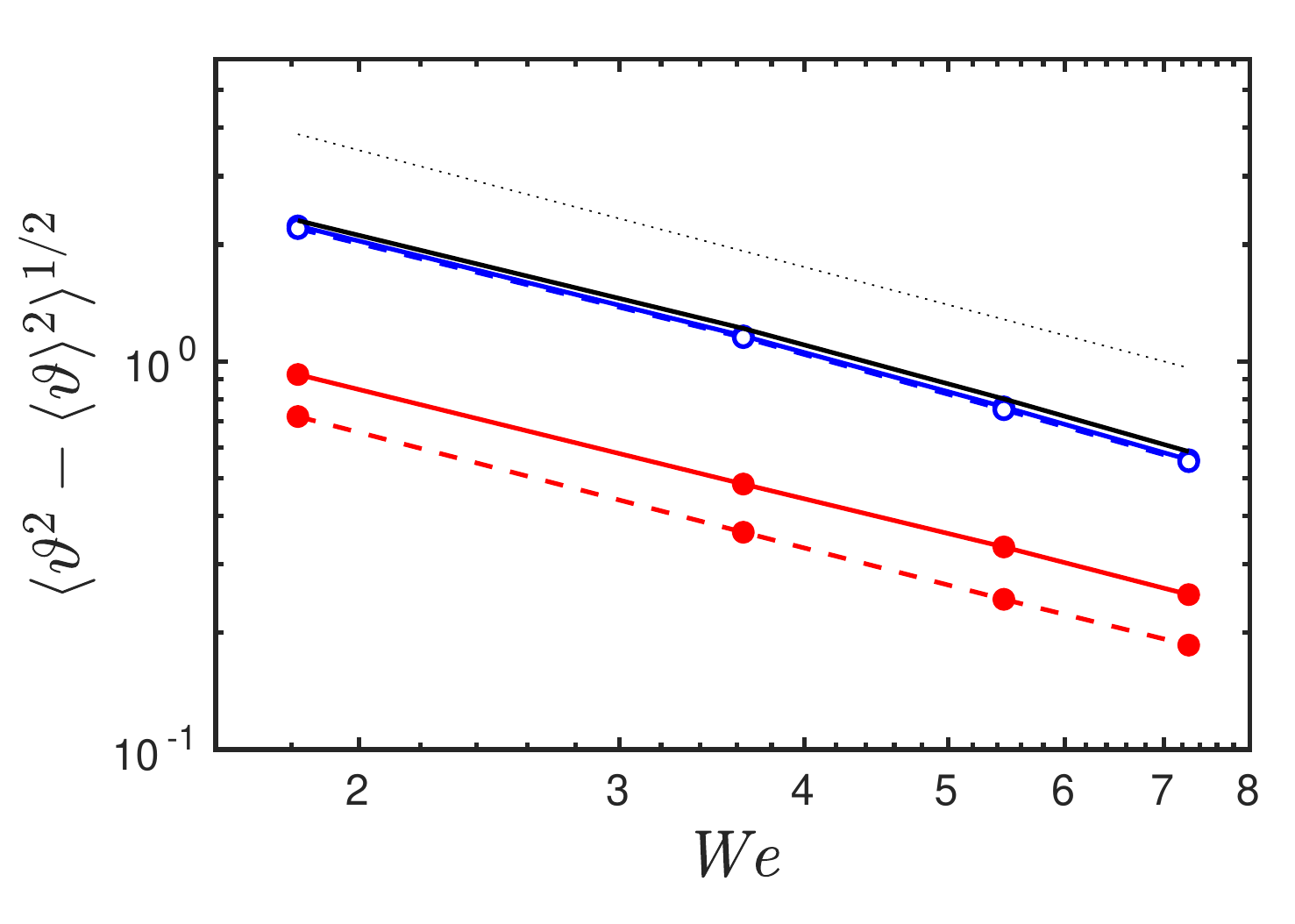}\\
(c)\vspace{-10pt}&
(d)\\
\includegraphics[width=0.5\textwidth]{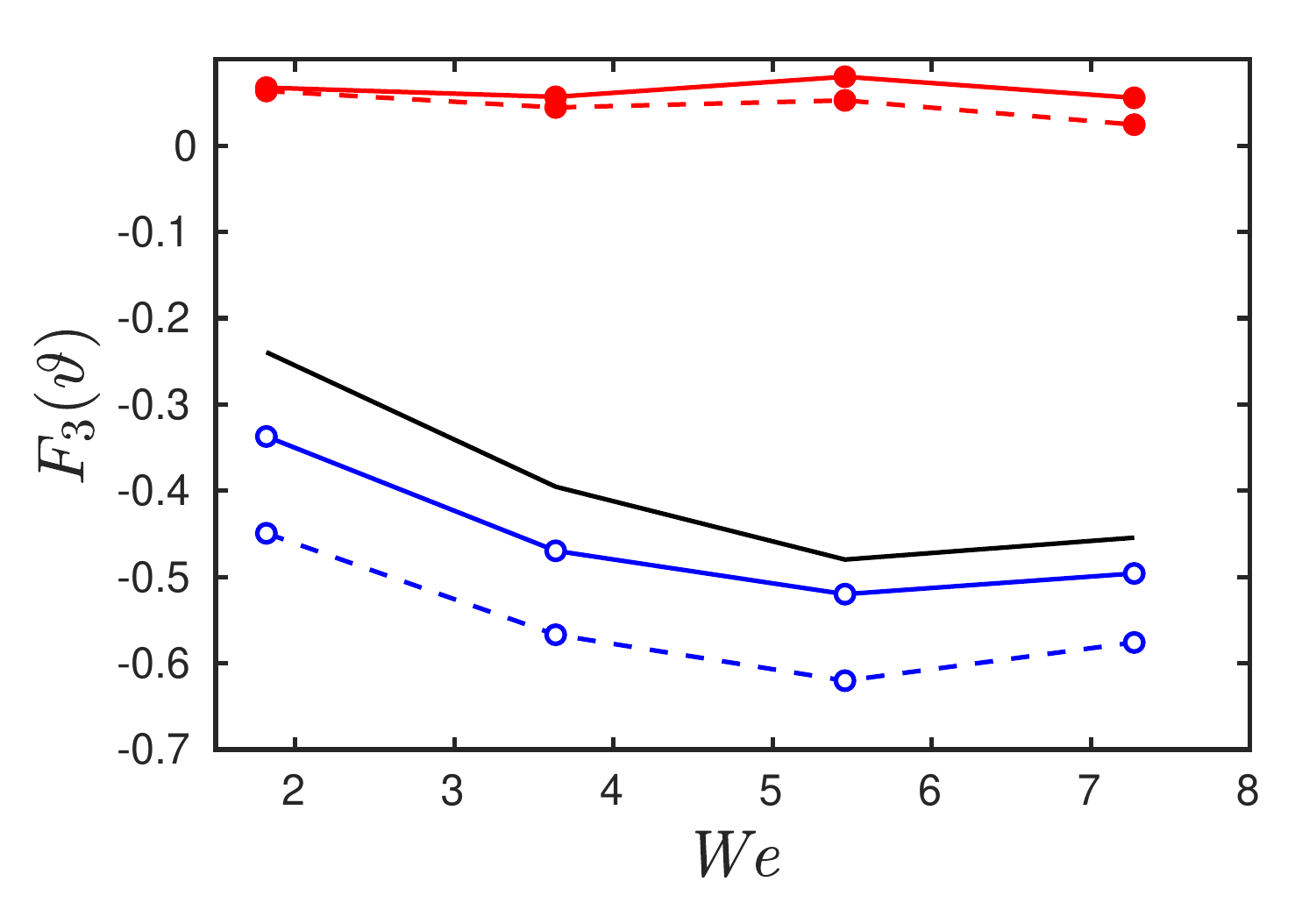}&
\includegraphics[width=0.5\textwidth]{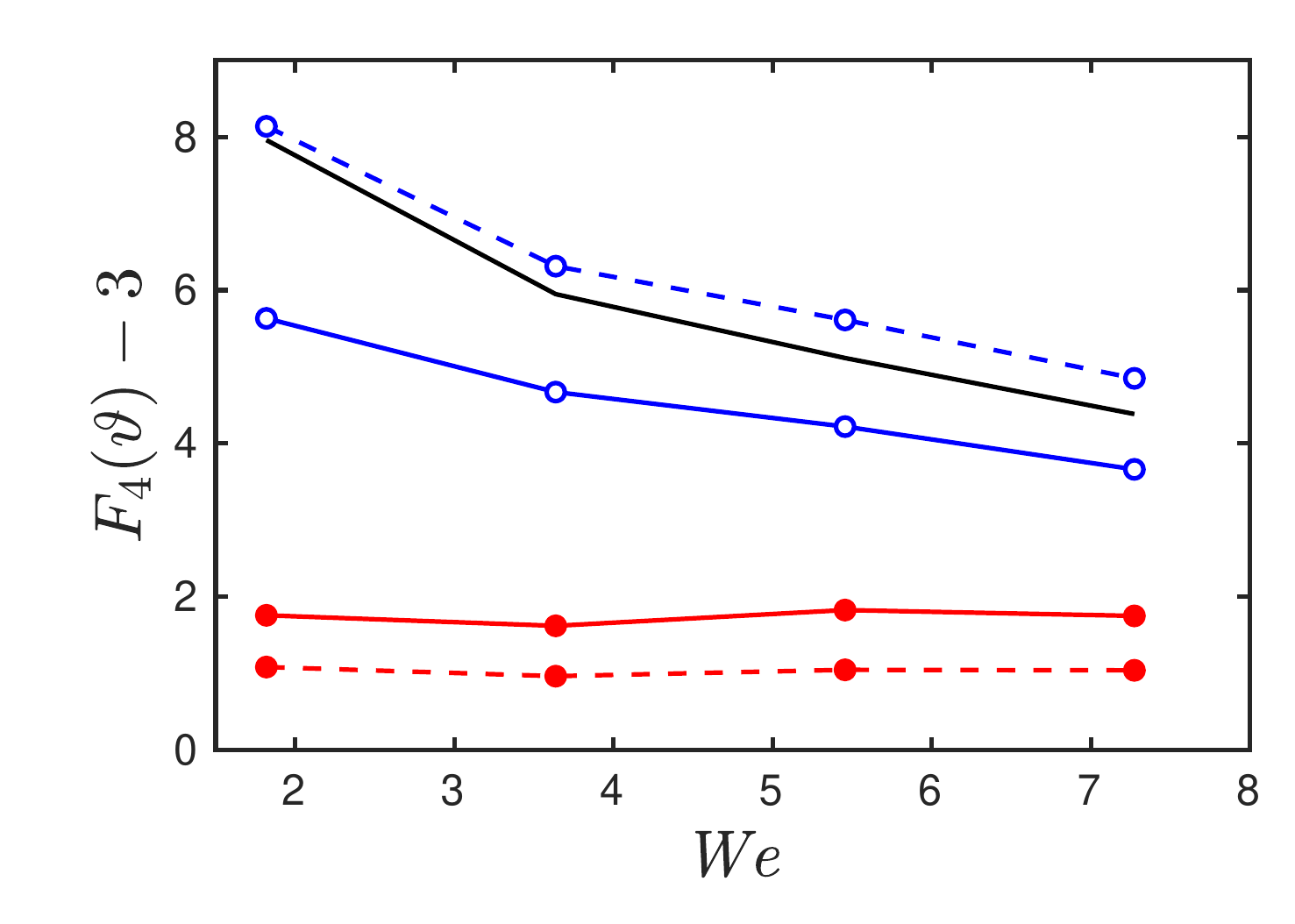}\\
\end{tabular}
\caption{(a) Mean, (b) standard deviation, (c) skewness and (d) excess flatness of $\vartheta$ as a function of $\We$. Mean and standard deviation normalised with $\rho u_d^3$, where $u_d=d/t_d$.
Solid symbols corresponds to the statistics of $\vartheta^O$, and empty symbols to $\vartheta^I$. Colour lines correspond to $\varDelta=$: 
\protect\solid, $0.12d$;
\protect\dashed, $0.19d$; 
and solid black line to the full rate-of-strain tensor.
In (b) the dotted line is proportional to $\We^{-1}$.}
\label{fig:mech2}
\end{figure}

In figure~\ref{fig:mech2}(a), we show the mean of $\vartheta$, $\vartheta^O$, and $\vartheta^I$ normalised with $\rho u^3_d$, 
where $u_d=d/t_d$, for different values of $\We$. The average of $\vartheta$ increases with $\We$, 
which is consistent with the statistics of the breakup time shown in figure \ref{spectra}(a). 
We assume that the breakup of an initially spherical drop
takes place when a critical increment of the surface energy, $\varDelta \mathcal{H}_b$, is reached \citep{andersson2006}. 
Because the surface energy can only increase due to the stretching term, its average measured on the surface of the drop is related to the time to breakup by
\begin{equation}
    t_b\sim\big\langle\frac{\varDelta \mathcal{H}_b}{\vartheta d^2}\big\rangle,
    \label{ref:33}
\end{equation}
which has been estimated by averaging equation (\ref{eq:ener31}) in time, considering that the increment of the surface energy is approximately 
$\langle\vartheta\rangle d^2$, and neglecting the dissipation of surface energy. 
\com{The increment of the surface energy required for breakup is approximately the difference between the surface energy after breakup, $\mathcal H_b$,
and the surface energy of a spherical drop, $\mathcal{H}_0$, i.e., $\varDelta\mathcal{H}_b\approx \mathcal{H}_b - \mathcal{H}_0=f\sigma d^2$,
where $f\leq f_{b}$ is a factor that depends on the geometry of breakup and is largest for binary breakup, when $f_{b}=2^{1/3} - 1\approx0.26$ \citep{andersson2006}.}
The time to breakup increases with decreasing $\We$, which is consistent with the increment of $\langle\vartheta\rangle$ with $\We$.

The average outer stretching is positive and larger than the inner stretching. It decreases approximately by a factor of $2$ from $\We=1.8$ to $\We=7.4$.
The inner stretching increases substantially with $\We$, transitioning from $\langle\vartheta^I\rangle<0$ to $\langle \vartheta^I\rangle>0$ at approximately $\We\approx3.5$. Note that negative average values of the inner stretching indicates that, close to the surface, energy is being mostly transferred from the surface to turbulent fluctuations. This phenomenon will be examined in detail in $\S$\ref{sub:44}.

%The procedure used to construct the database, in which simulations are stopped when the drop breaks, implies that $\langle \varDelta \mathcal H/t_b\rangle\sim\langle\vartheta\rangle$, where $\varDelta \mathcal H$ is the surface energy increment necesary to break the drop, and $t_b$ aasdfis the time to breakup.
%The fact that $\langle\vartheta\rangle$ increases %with $\We$
%reflects that, as expected, the average time to %breakup decreases with increasing $\We$. 

%\textcolor{red}{Marc: should we comment on the implications or meaning of this scaling (or is it done later)?}
%This scaling is very clear for the outer contributions, %whereas, 
%for the outer contributions, we observe some deviation from this scaling at low $\We$.
% Marc: I have commented this statement out because we don't refer to each later and it does not really add valuable information, I believe

In figure \ref{fig:mech2}(b), we show the standard deviation of $\vartheta$ normalised with $\rho u^3_d$.
It is approximately proportional to $\We^{-1}$, and substantially higher for the inner than for the outer stretching.
%%%%
In all cases, the standard deviation of the surface stretching is larger than its mean.
This is particularly significant for the inner stretching and small $\We$,
and suggests that a very significant part of the stretching is not efficient and cancels out when averaging. 

We focus now on higher-order statistics of the surface stretching, in particular the skewness and the flatness factor, which are defined as
\begin{equation}
     F_n(\vartheta)=\frac{\langle(\vartheta-\langle\vartheta\rangle)^n\rangle}{\langle\vartheta^2 -\langle\vartheta\rangle^2\rangle^{n/2}},
\end{equation}
%%%
for $n=3$, and $n=4$ respectively, and are shown in figures~\ref{fig:mech2}(c,d).
For ease of comparison, we consider the excess flatness factor, which is the flatness factor minus that of a Gaussian distribution, for which $F_4=3$. 
These statistical moments reflect the strong differences between inner and outer stretching.
While the skewness of the outer contributions is close to zero and does not change with $\We$, the inner stretching is negatively skewed,
and its skewness depends on the $\We$. The excess flatness factor conveys a similar picture. 
The outer contributions have a low, $\We$-independent, excess flatness factor,
whereas for the inner contributions the excess flatness factor increases with decreasing $\We$.
These results corroborate, first, that the statistics of the outer stretching are $\We$-independent,
and, second, that the inner stretching has an intermittent structure, with intense events of the surface stretching becoming stronger with decreasing $\We$. These events are probably related to the relaxation of the drop towards a spherical shape.

\begin{figure}
\centering
\begin{tabular}{ll}
(a)&
(b)\\
\includegraphics[width=0.48\textwidth]{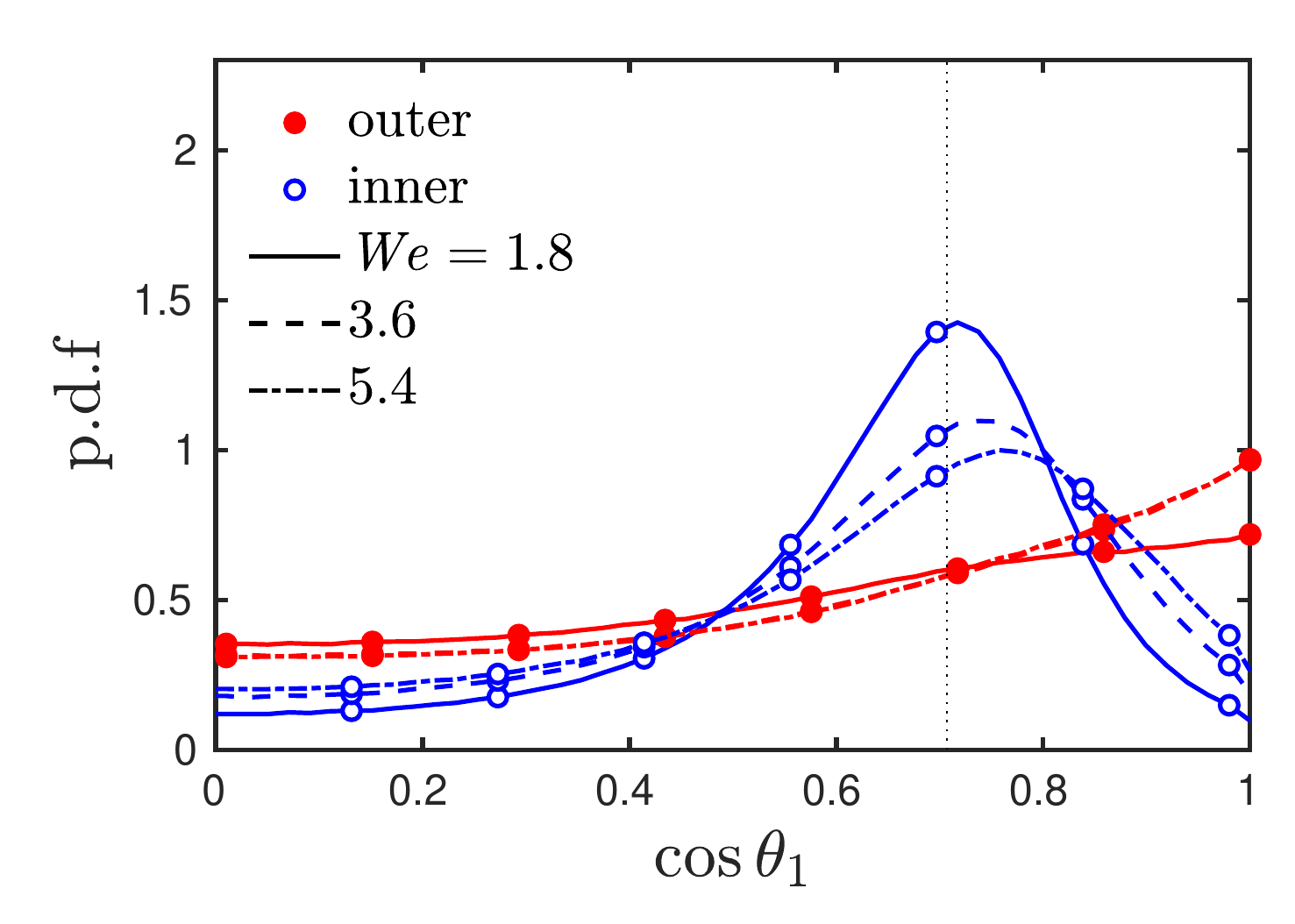}&
\includegraphics[width=0.48\textwidth]{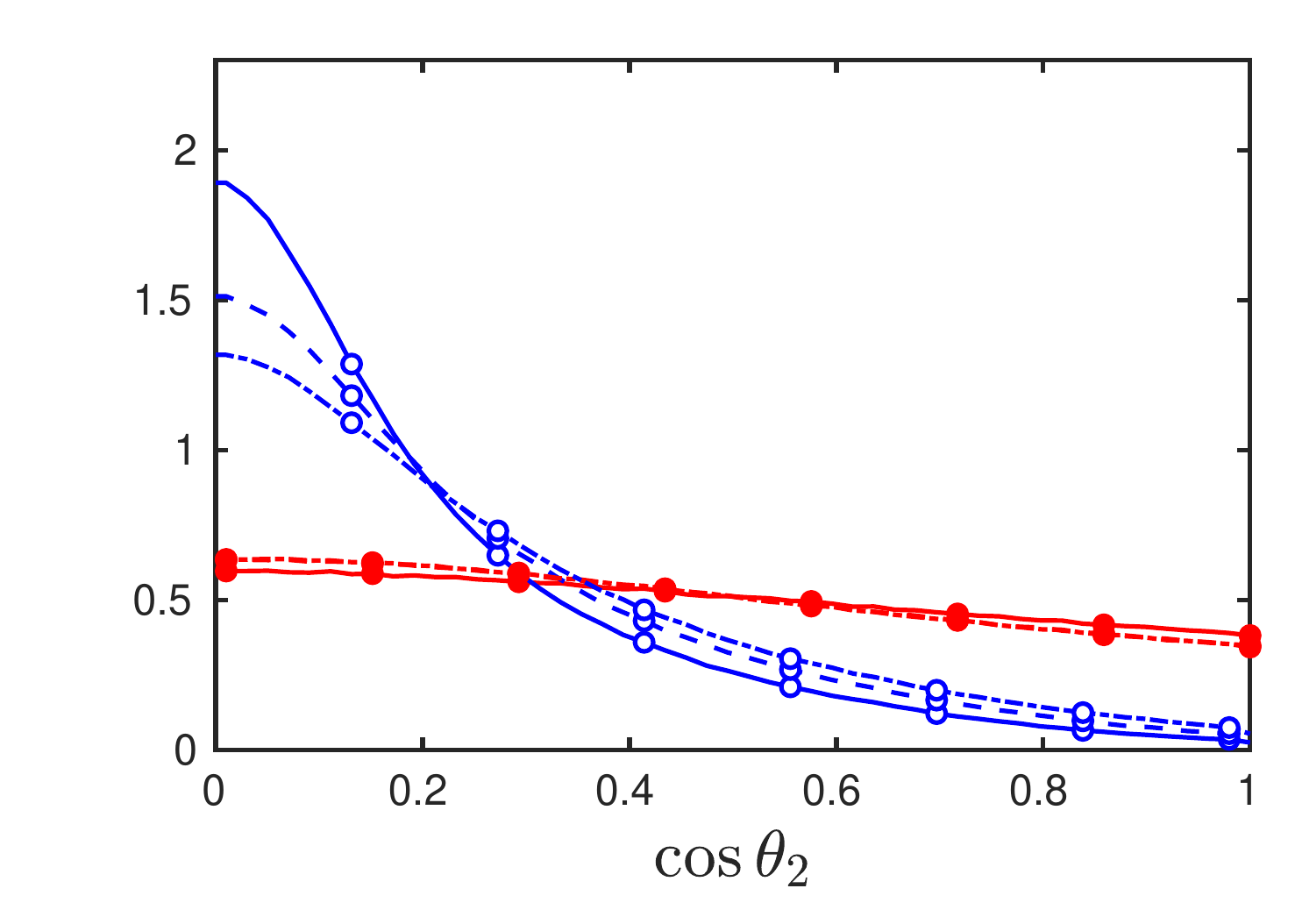}\\
(c)\vspace{-10pt}&
(d)\\
\includegraphics[width=0.48\textwidth]{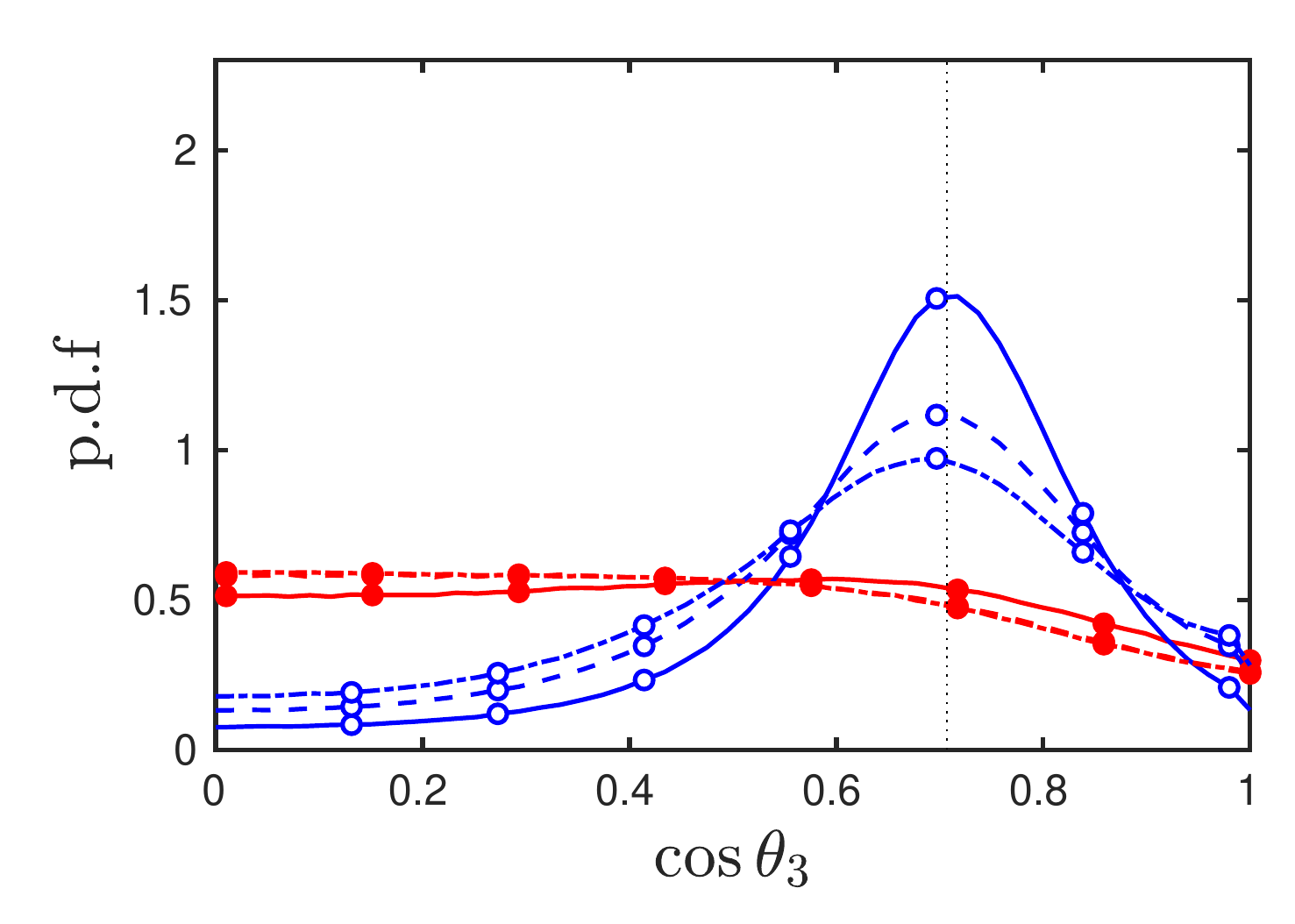}&
\includegraphics[width=0.48\textwidth]{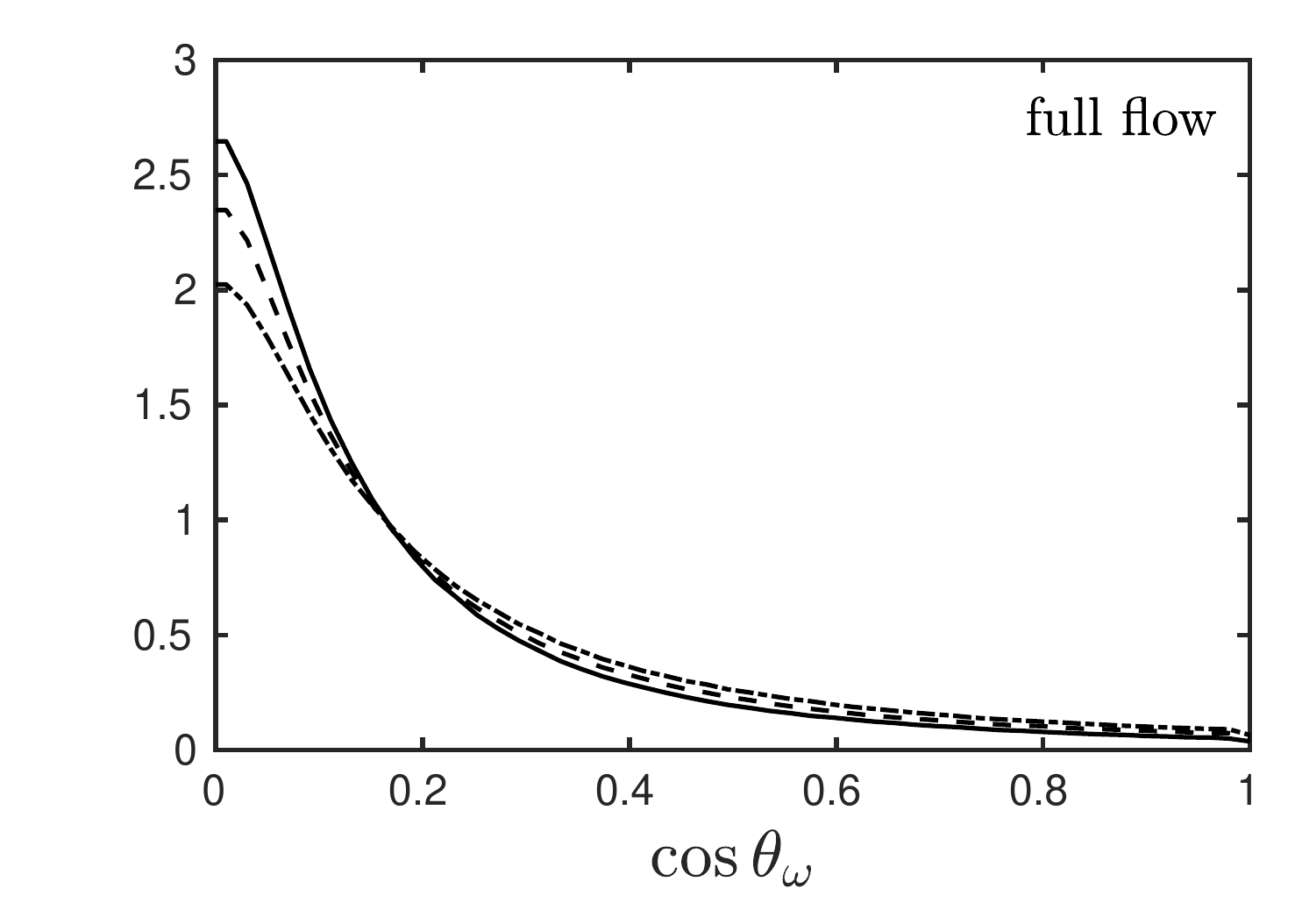}\\
\end{tabular}
\caption{
(a)-(c) Probability density function of $\cos \theta_i=\boldsymbol n \cdot \boldsymbol v_i$, where $\boldsymbol v_i$ are the principal directions of the rate of strain tensor and $\lambda_1\le\lambda_2\le\lambda_3$ are their eigenvalues.
Solid markers correspond to angles calculated with $S^{O}_{ij}$, and empty markers to $S^{I}_{ij}$, for $\varDelta=0.19d$. 
(d) Similar but for the vorticity vector, $\cos \theta_\omega=\boldsymbol n \cdot \boldsymbol \omega/|\boldsymbol \omega|$.
Lines correspond to $\mathrm{\it We}=$:
\protect\solid, $1.8$;
\protect\dashed, $3.6$;
\protect\dotdashed, $5.4$.
The vertical dotted line in (a) and (c) marks $\cos \pi/4$.
}
\label{fig:mech3}
\end{figure}

\subsection{Geometrical characterisation of the inner and outer stretching}

To further explain the differences between the inner and outer stretching, and how they change with $\We$,
we study the structure of the vorticity vector and the rate-of-strain tensor induced by inner and outer eddies on the surface of the drop. 
In the spirit of the analysis of vortex stretching in isotropic turbulence \citep{ashurst1987alignment,buaria2020vortex},
we analyse the surface stretching term in the frame of reference 
of the eigenvectors of the rate-of-strain tensor,
and quantify the contribution of each of its eigenvalues to the total surface stretching.

In figure~\ref{fig:mech3}(a)--(c), %(a-d),  
we show the p.d.f of the cosine of the angle of alignment between each of the principal axes of the rate-of-strain tensor, $\bs v_1$, $\bs v_2$ and $\bs v_3$ (where $\lambda_1\le\lambda_2\le\lambda_3$ are their corresponding eigenvalues) and the normal to the surface, $\boldsymbol n$. 
%We have separatedly performed this decomposition for rate-of-strain tensor induced on the surface of the drop by inner and outer eddies.
% 
% The rate-of-strain tensor and the vorticity induced by eddies within 
% $\varDelta=9\eta$ of the surface are very affected by the interface.
% 
%
For the inner contributions, the most stretching ($\boldsymbol{v}_3$) and the most compressing ($\boldsymbol{v}_1$) eigenvectors tend to be oriented at $\sim45^o$ with respect to the surface normal, and therefore also to the surface tangent plane, while the intermediate eigenvalue is predominantly normal to the surface normal (parallel to the surface tangent plane). 
As shown in figure \ref{fig:mech3}(d), also the vorticity vector aligns strongly normal to $\boldsymbol n$, and parallel to the surface. This trend was also reported by \citet{soligo2020effect}. In all cases, the alignment is more marked for small $\We$, revealing an important effect of surface dynamics on the configuration of the velocity gradients at the surface of the drop.
%We note that the surface cannot generate vorticity normal to it due to $\boldsymbol n \cdot\nabla\times(\phi\nabla c)=0$. 

\begin{figure}
\centering
\begin{tabular}{ll}
(a)&
(b)\\
\includegraphics[width=0.48\textwidth]{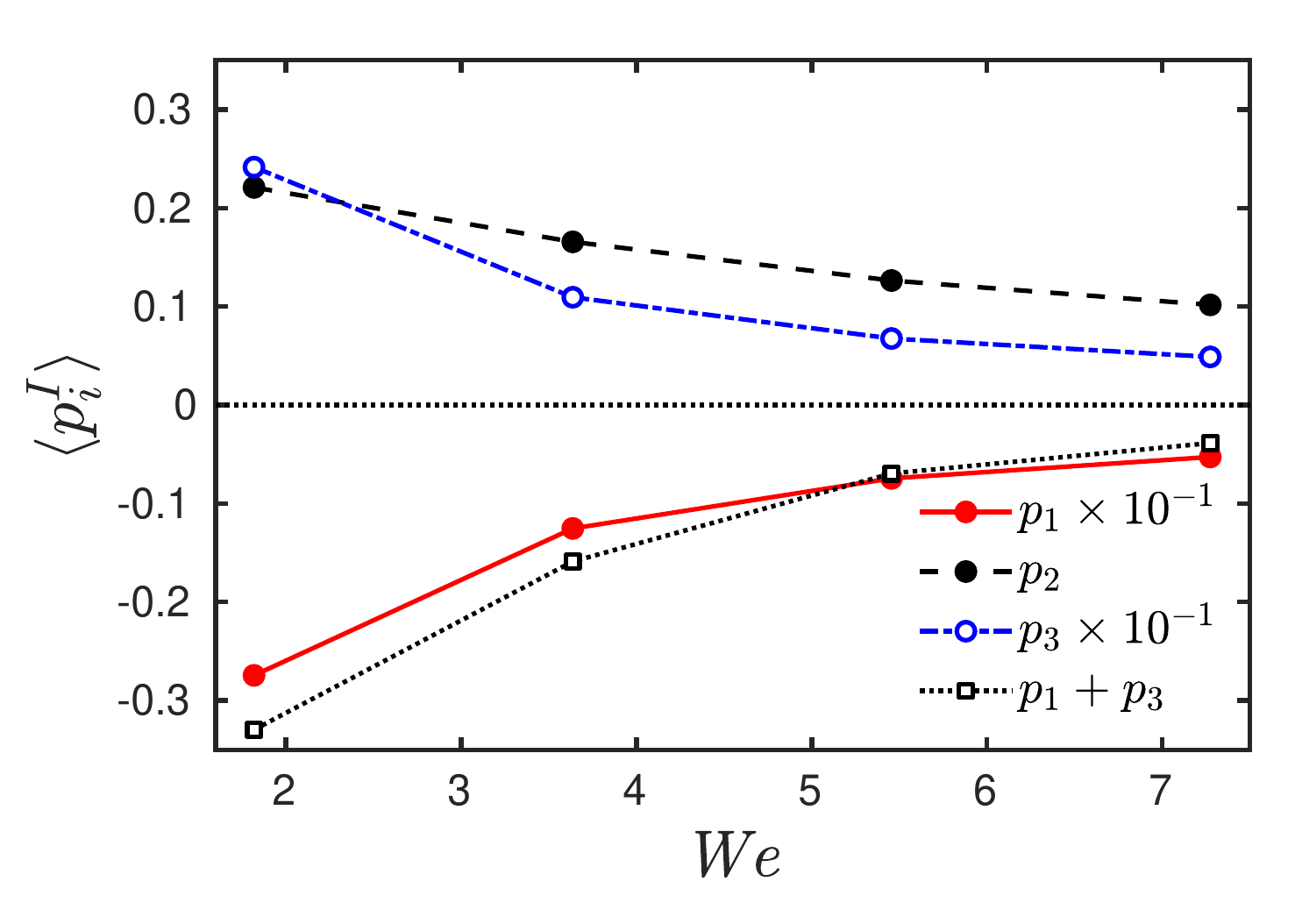}&
\includegraphics[width=0.48\textwidth]{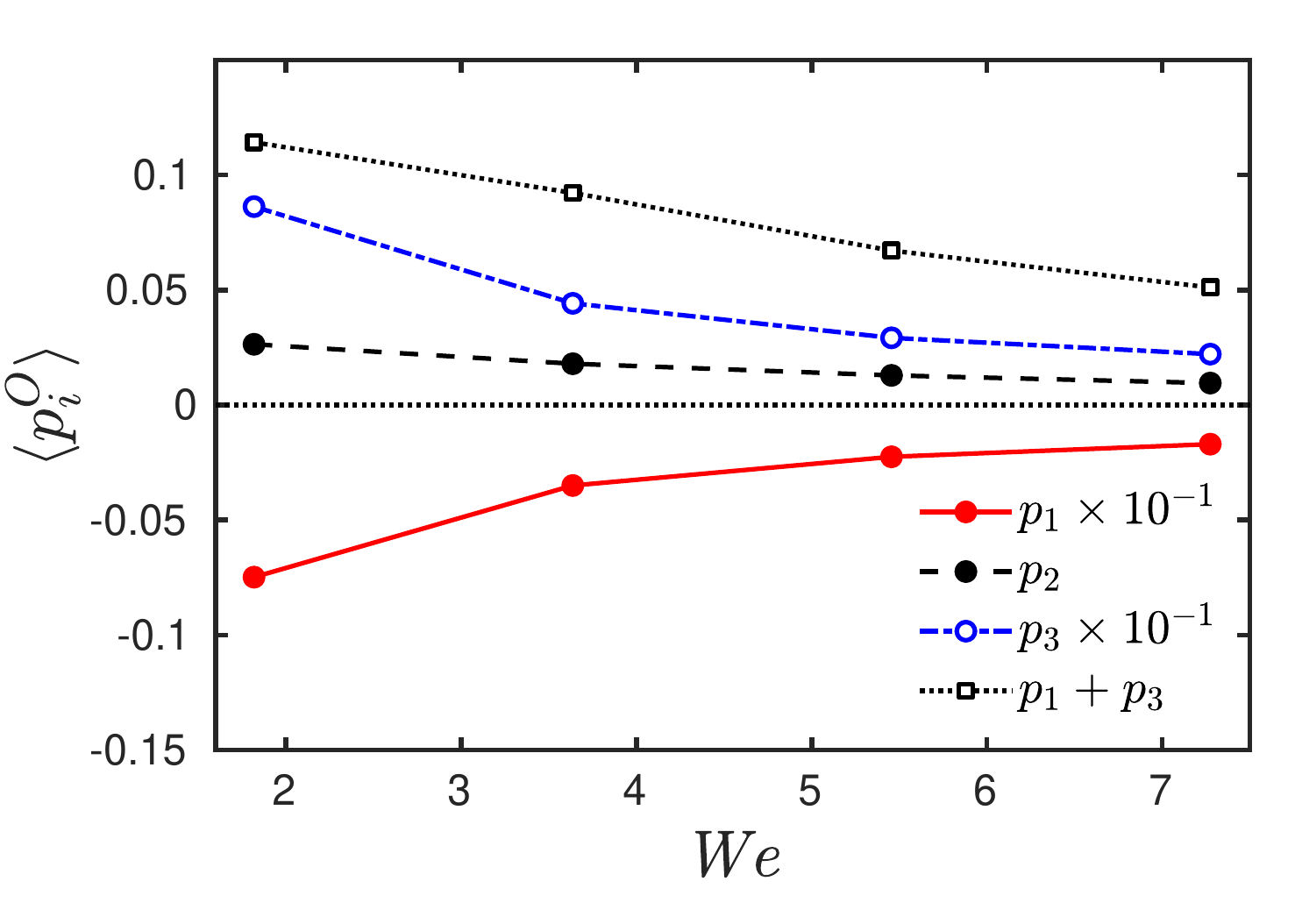}\\
(c)\vspace{-10pt}&
(d)\\
\includegraphics[width=0.48\textwidth]{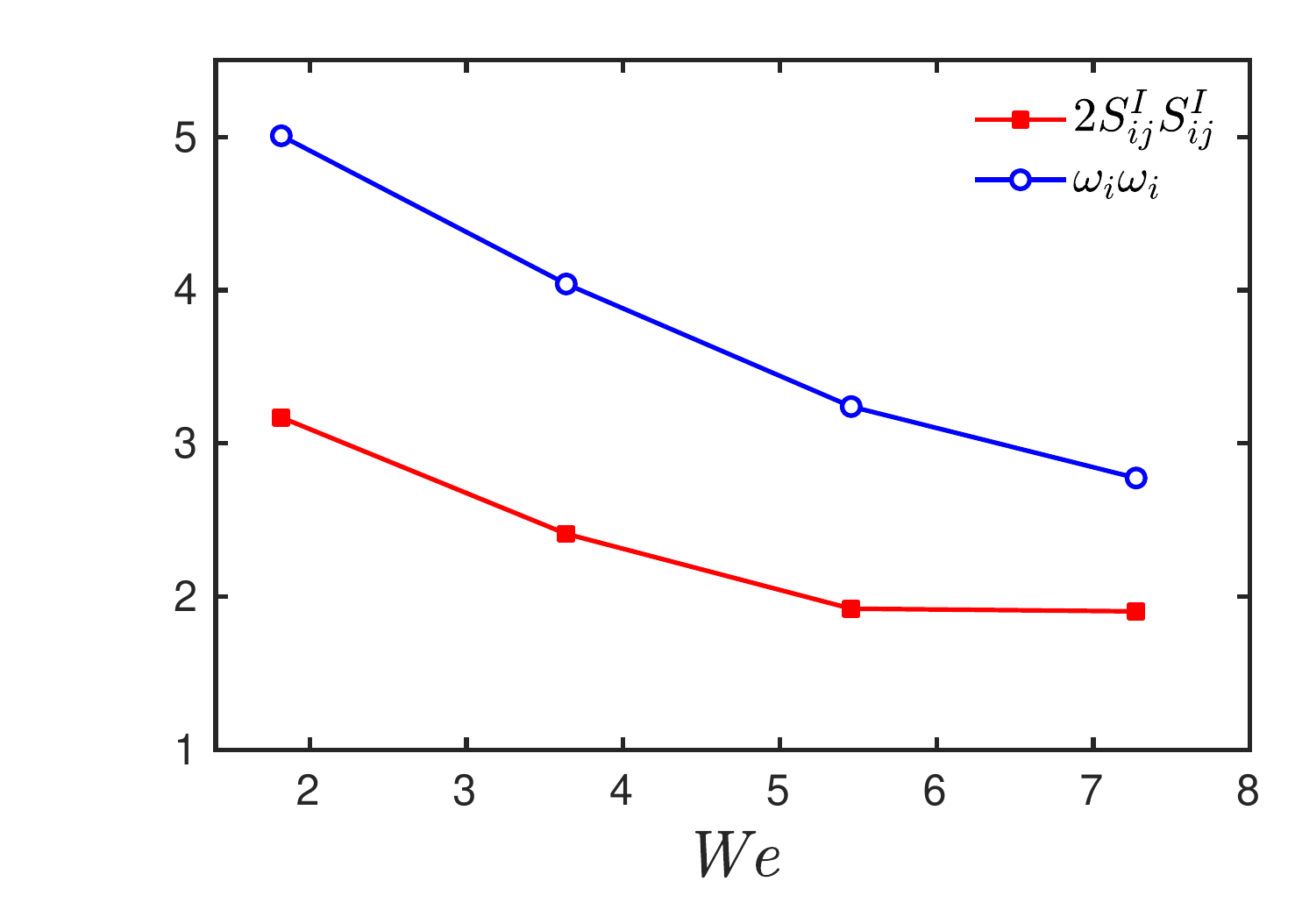}&
\includegraphics[width=0.48\textwidth]{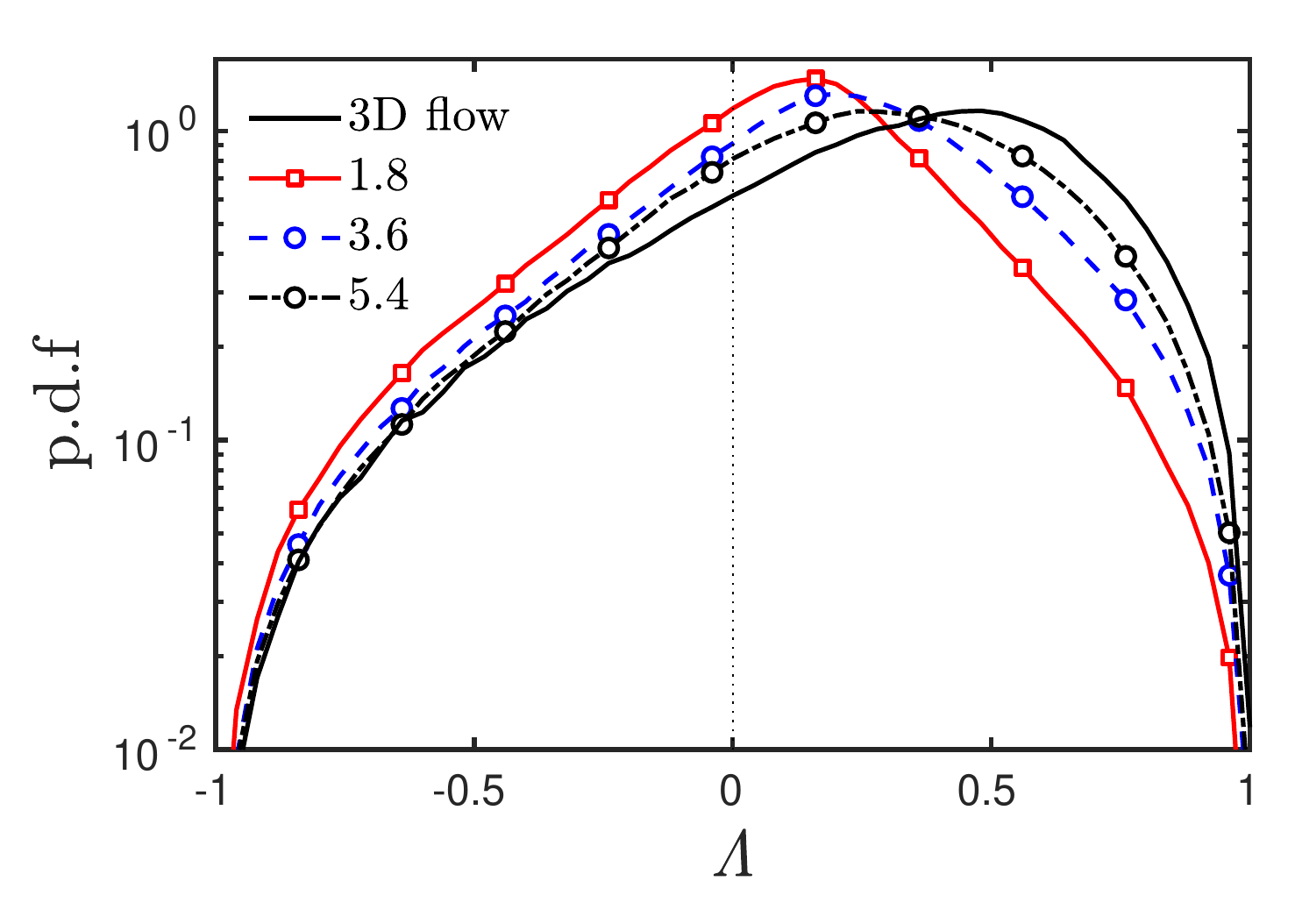}\\
\end{tabular}
\caption{(a,b) Average surface stretching due to each eigenvalue of the rate-of-strain tensor, $p_i=\sigma\lambda_i\sin^2\theta_i$ as a function of the Weber number for the inner (a) and the outer (b) contributions, for $\varDelta=0.19d$. The surface stretching is normalised with $\rho u_d^3$.
(c) Surface average of the square of the inner  rate-of-strain tensor, $2S^I_{ij}S^I_{ij}$, for $\varDelta=0.19d\eta$,
and the vorticity vector $\omega_i\omega_i$ for different Weber numbers, normalised with Kolmogorov units.
(d) p.d.f of $\Lambda=\log_2 |\lambda_1|/\lambda_3$ for different Weber numbers.}
\label{fig:mech4}
\end{figure}

The stretching of the surface by outer eddies shows a substantially different picture. 
There is a weak but consistent tendency of the most compressing eigenvector to align normal to the surface tangent plane. % although the alignment is less marked than for the inner rate-of-strain tensor.
The collapse of the p.d.f's of $\cos \theta_i$ at different $\We$ suggests that this effect is independent of the surface dynamics.

We further decompose the surface stretching term into the contribution of each of its eigenvalues.
A possible approach is to consider the rate of compression or stretching of the vector normal to the surface by each eigenvalue, $r_i=-\sigma\lambda_i\sin\theta_i $ (no summation for repeated indices is intended), which yields 
\begin{equation}
    \vartheta=r_1 + r_2 + r_3=-\sigma(\lambda_1\cos^2\theta_1 + \lambda_2\cos^2\theta_2 + \lambda_3\cos^2\theta_3).
    \label{111}
\end{equation}
However this decomposition cannot be readily interpreted from a physical perspective. Instead, we consider the contribution of each eigenvalue to the stretching or compression of the surface tangent plane, $p_i=\sigma\lambda_i\sin^2\theta_i$.
Note that $\sin\theta_i=\cos(\pi/2-\theta_i)$ is the projection of each eigenvector on the surface tangent plane.
Since $\lambda_1 + \lambda_2 + \lambda_3=0$, and $\cos^2\theta_i=1-\sin^2\theta_i$, the total surface stretching reads,
\begin{equation}
    \vartheta=p_1 + p_2 + p_3=\sigma(\lambda_1\sin^2\theta_1 + \lambda_2\sin^2\theta_2 + \lambda_3\sin^2\theta_3).
    \label{222}
\end{equation}
%%

%CAMBIAR EL ORDER DEL ANALYSIS
%To give this decomposition a physical meaning, we transform the compression/stretching of the surface-normal vector into the 
%%
%\begin{equation}
%    \vartheta=\sigma(\lambda_1\sin^2\theta_1 + \lambda_2\sin^2\theta_2 + \lambda_3\sin^2\theta_3),
%\end{equation}
%%
%%
%where $p_i=\sigma\lambda_i\sin^2\theta_i$ is interpreted as the contribution of each eigenvalue to the stretching or compression of the surface tangent plane.
%Note that $\sin\theta_i=\cos\theta_i^s$, where $\theta^s_i=\pi/2-\theta_i$ is the angle between each eigenvector and the surface tangent plane, and  $p_i=\sigma\lambda_i\cos^2\theta^s_i$, where $\cos\theta^s_i$ represents the projection of each eigenvector on the surface tangent plane.
%INCLUDE THE BRACKETS, TALK ABOUT MEANS

In figures \ref{fig:mech4}(a,b), we show the averages of $p_i$ for the inner and outer contributions. In both cases the surface stretching and compression due to $p_1$ and $p_3$ is much larger in absolute value than that of $p_2$ (note that in the plot $p_1$ and $p_3$ are divided by $10$ for ease of visualisation).
However, there is a significant cancellation between $p_1$ and $p_3$, and $\langle p_1 + p_3\rangle\sim\langle p_2\rangle$.

For the outer stretching, $p_1 + p_3$ produce net surface stretching, which stems from the tendency of $\boldsymbol v_3$ to align parallel to the surface (or $\boldsymbol v_1$ normal to it). Conversely, $p_2$ does not contribute on average to the outer stretching.
This is in agreement with the almost random orientation of $\boldsymbol v_2$ with respect to the surface normal, as shown in figure \ref{fig:mech3}(b).

For the inner stretching, the cancellation of $p_1$ and $p_3$ is consistent with a predominant 
alignment of $\boldsymbol v_1$ and $\boldsymbol v_3$ at $\sim 45^o$ with respect to the normal
and to the surface tangent plane. 
In this case, $p_2$ contributes to the average stretching of the surface. 
As $\We$ increases, the absolute value of $\langle p_1 + p_3\rangle$ decreases,
and $\langle p_2 \rangle$ progressively dominates the average stretching, 
leading to $\langle \vartheta^I\rangle>0$ for $\We>3.5$, as shown in figure \ref{fig:mech2}(a).

%SUBSECCION: TRANSITION FROM 2D 
\subsection{Transition from a $2D$ to a $3D$ structure the inner rate-of-strain tensor}
\label{sub:44}

In this section, we explain the transition from $\langle \vartheta^I\rangle<0$ to $\langle \vartheta^I\rangle>0$ at $\We\approx3.5$
by analysing the intensity and structure of the velocity gradients at the drop surface. 
In figure \ref{fig:mech4}(c), we show the average  of $2S^I_{ij}S^I_{ij}$ for $\varDelta=0.19d$, and $\omega_i\omega_i$, measured at the surface of the drop.
The intensity of the velocity gradients on the surface of the drop increase with decreasing $\We$ due to surface tension forces,
but these forces impose a strong constraint on their structure.
% but these forces also constraint their structure.
% Although intense, the velocity gradients are strongly constrained .
%which precludes the stretching of the surface. This constraint, however, does not affect the dissipation of turbulent kinetic energy, which is enhanced close to the drop surface, specially for low $\We$.
%which explains%the transition of $\langle\vartheta^I\rangle$.
%
This is particularly noticeable in the rate-of-strain tensor, which has a quasi-2D structure at the low $\We$, with $|\lambda_2|\ll|\lambda_3|\approx|\lambda_1|$, and becomes progressively 3D, $|\lambda_2|\sim|\lambda_3|\sim|\lambda_1|$, as $\We$ increases. 
To characterise this transition we analyse
\begin{equation}
\Lambda=\log_2 \frac{|\lambda_1|}{\lambda_3},
\end{equation}
at the drop surface. Since $\tfrac{1}{2}\lambda_3\leq|\lambda_1|\leq2\lambda_3$, this quantity is defined  in the range $-1\leq\Lambda\leq1$.
Predominant $\Lambda>0$ is characteristic of fully developed 3D turbulence, and implies that $\lambda_2>0$ on average.
Conversely, $\Lambda=0$ indicates 2D dynamics, in which $\lambda_1=-\lambda_3$ and, by continuity, $\lambda_2=-\lambda_1 - \lambda_3=0$.
As shown in figure \ref{fig:mech4}(d), the p.d.f of $\Lambda$
shows that, for small $\We$, $|\lambda_1|\sim\lambda_3$ and $\lambda_2\ll\lambda_3$. 
As $\We$ increases, the peak of the p.d.f of $\Lambda$ moves towards $\Lambda\approx0.4$,
indicating a transition from a quasi-2D to a 3D structure of the inner rate-of-strain tensor at the surface. 
This transition implies that $\lambda_3$ and $|\lambda_1|$ decrease with respect to $\lambda_2$,
which leads to $\langle p_2\rangle>\langle p_1 + p_3 \rangle$, and to $\langle\vartheta^I\rangle>0$.

%In fact, the alignment angles of $\bs v_1$ and $\bs v_3$ with the normal at $\sim45^o$ are consistent with the rate-of-strain field generated by strong vortices whose axis is contained in the plane tangent to the surface \citep{jimenez1992kinematic}.
%\section{Discussion}

\section{Discussion}
\label{sec:diss}

We have separated the stretching of the drop surface in contributions from inner and outer eddies, which we have defined in $\S$\ref{sec:decomp2}. 
For distances to the drop surface larger than $\varDelta=0.12d$, the inner and outer stretching are not correlated pointwise 
(figure \ref{fig:mech1}), indicating that beyond this distance the outer stretching is independent of surface dynamics. 
This is corroborated by the statistics of the outer surface stretching for this $\varDelta$, which are roughly $\We$-independent.
We have shown that the outer stretching is dominant and contributes on average to the increment of the surface energy at all $\We$. 

Conversely, the average of the inner stretching is negative for $\We<3.5$, 
indicating a predominant flux of energy from the surface into turbulent fluctuations.
We interpret this flux of energy as the generation of inner eddies due to the relaxation of the surface.
Above $\We\approx3.5$, the average inner stretching becomes positive, indicating that, on average, also inner eddies contribute to drop deformation.
In all cases, the inner stretching shows non-Gaussian and negatively-skewed $\We$-dependent statistics. 
Negative events of the inner stretching become more intense as $\We$ decreases, 
evidencing the action of the surface to restore deformations by transferring the surface energy back to turbulent fluctuations.

At low $\We$, the net increments of the total surface energy depend on the interplay between the outer surface stretching,
which constitutes a mechanism of net drop deformation, and the ability of the surface to restore this deformation 
by transferring energy to inner eddies, which provide a source of energy dissipation through molecular viscosity.
In addition, we have shown that at low $\We$ the rate-of-strain tensor is constrained by surface tension forces.
Although its average magnitude is larger than in the turbulent background,
providing enhanced dissipation of turbulent kinetic energy, 
its 2D structure and its configuration with respect to the interface precludes the net stretching of the surface. 
As $\We$ increases, the inner rate-of-strain tensor develops a 3D structure in which its intermediate eigenvalue 
becomes predominantly positive, producing net stretching of the surface.
This phenomenon explains the transition at $\We\approx3.5$ from a net energy transfer from the surface to inner eddies (negative inner stretching), 
to a net energy transfer from inner eddies to the surface (positive inner stretching). 

We have reported a statistically significant alignment of the surface normal vector
with the most compressing direction of the rate-of-strain tensor induced by outer eddies,
and of the surface tangent plane with the most stretching one.
This phenomenon, which is also observed in the evolution of passive material surfaces in turbulence, 
and naturally leads to these alignments \citep{girimaji1990material},
indicates the persistent stretching of the drop surface by outer eddies.
The alignment statistics are similar at different $\We$, corroborating
that this mechanism is independent of surface dynamics.

\section{Conclusions}
\label{sec:con}

%The physical interpretation of the exchange mechanism is intuitive. The surface energy of an interface 
%is proportional to its surface, and therefore changes proportional to it. This is precisely what the energetic exchange term describes as the stretching or compression of the vector normal to the surface, in a mechanism that resembles the stretching of vorticity lines by the rate-of-strain tensor. In incompressible flows, the compression/stretching of the surface normal is equivalent to the stretching/compression of the surface area. A similar interpretation is possible for compressible flows, but including an extra term for the local volume expansion or contraction. 
%Our analysis points to the central role of the rate-of-strain tensor in the evolution of the surface energy in fluid-fluid interfaces.
%By applying mathematical analysis to the Cahn--Hilliard--Navier--Stokes (CHNS) equations \citep{jacqmin1999calculation}, we have proved that this exchange occurs only due to the stretching or compression of the surface area by the rate-of-strain tensor. Although our derivation benefits from the use of a diffuse interface method, the consistency of the CHNS formulation in the limit of the zero-thickness interface \citep{magaletti2013sharp} 
%ensures that our result is physically universal. 
%In fact, we have shown that it holds regardless of the viscosity and density ratios of the carrier and disperse phase, and is also valid for compressible flows as long as the velocity field does not contain any discontinuity.

In this work we have analysed the energetics of fluid-fluid interfaces. 
One of our main contributions is to provide a mathematical description of the mechanism
responsible for the exchange between the kinetic energy of the flow
and the surface energy of the fluid-fluid interface.
We have shown that this exchange occurs due to the stretching 
or compression of the surface area by the rate-of-strain tensor 
(see (\ref{eq:stretching})-(\ref{eq:ener322})),
in a mechanism analogous to the stretching of the vorticity field in turbulence.
This analytical result highlights the relevant role of straining motions 
in drop deformation and breakup, in agreement with experiments and theoretical analyses of laminar flows in the Stokes limit \citep{rallison1984}.

In the context of drop deformation and breakup, this mathematical analysis provides a solid framework 
to quantify the energetic exchanges between the surface and the kinetic energy fluctuations of the surrounding turbulence.
By leveraging the non-local nature of turbulence,
we have separated the total stretching into contributions due to inner and outer eddies. 
We use here the term outer to refer to eddies sufficiently far from the drop surface 
for their dynamics to be unaffected by surface tension forces. 
Conversely, we define inner eddies as those close to the surface or inside the drop, 
whose dynamics are affected either by surface tension forces or by the material properties of the fluid inside the drop.  
We have shown that the stretching of the drop surface by outer eddies contributes substantially to the deformation of the drop,
and is relevant to drop breakup. This mechanism constitutes a physically well-defined, quantitative reinterpretation 
of the phenomenological `collision' of eddies \citep{luo1996}. 
% On the other hand, for $\We<3.5$ inner eddies 

% Thus their contribution to surface stretching may be considered separately. 

Our results justify a `stochastic' approach to breakup modelling, particularly at low $\We$. 
Outer eddies provide the main source of drop deformation.
They are not locally coupled to inner eddies or to the dynamics of the interface, which precludes synergies between inner and outer eddies
that may lead to enhanced breakup
(at least in the case of similar density and viscosity investigated here).
For sufficiently long breakup times, the stretching by outer eddies may be modelled as 
a stochastic forcing with statistical properties that depend mostly on the surrounding turbulence.
\com{Although the surface stretching by inner and outer eddies is not coupled locally, it is likely globally coupled, for instance, 
due to energy conservation.}

Our results suggest that for low $\We$ ($\We<3.5$) drop deformation and breakup occurs as an interplay between inner and outer dynamics, 
where the latter is the driver and a source of surface energy increments, while the former is a sink.
The stretching by outer eddies is mostly balanced by an energy transfer from the surface to inner eddies,
which provide a source of dissipation and prevent the buildup of surface energy.

% Another source of coupling comes from the geometry of the drop,
% which is driven by, and at the same time affects, the stretching by inner and outer eddies.
% In summary, drop deformation and breakup occurs as an interplay between inner and outer dynamics,
% where the latter is a source of surface energy increments while the former is a sink.

Although our study is limited to the ideal case of incompressible fluids with 
equal density and viscosity, the theoretical analysis in $\S$\ref{sec:anal} 
applies equally to compressible fluids, and to fluid pairs with different 
viscosity and density.
% we expect the contribution of inner dynamics to change considerably.
For instance, we suggest that an increment of the drop viscosity affects the inner eddies due, first,
to the kinematic relations imposed on the rate-of-strain tensor across the interface, 
which depend on the viscosity ratio \citep{dopazo2000vorticity},
and second, by the enhanced ability of the fluid inside the drop to diffuse momentum and dissipate energy. 
These changes may inhibit the stretching of the surface by inner eddies and/or 
provide a fast mechanism to dissipate the surface energy produced by the stretching of outer eddies,
explaining the resistance of viscous drops to breakup \citep{calabrese1986drop,roccon2017viscosity}.
Our analysis also applies to bubbles. We suggest that our 
decomposition would allows to separate the effect of the bubble's natural oscillations from that of turbulence,
allowing to explore how resonances with the characteristic oscillatory frequency affect breakup \citep{risso1998}.

% In bo, we suggest that outer contributions should remain similar.
% We suggest this may be a consequence of the lack of a strong dissipative mechanism for inner eddies, 
% which   

%Future work will be directed towards understanding the causal relations that operate between inner and outer dynamics.

An important question that we have not addressed in this work is whether breakup occurs as a progressive buildup of surface energy 
due to the accumulative interaction with many eddies, 
or as sharp increments of the surface energy due to interactions with isolated intense turbulent events.
Although answering this question necessarily requires temporal analysis,
we have shown that the standard deviation of the local surface stretching is much larger than its mean, 
possibly reflecting inefficient stretching that cancels out. 
Moreover, at low $\We$, the stretching by outer eddies is mostly balanced by a transfer of energy
back to turbulent fluctuations (inner eddies), precluding the buildup of surface energy.
These results suggest that the weak turbulent background produces much of this ineffective stretching, 
whereas the interaction of the drop with just a few intense turbulent structures
produces the effective stretching leading to breakup.

\section*{Acknowledgements}
A.V.-M. acknowledges the support of the European Research Council COTURB project ERC-2014.AdG-669505. 
We thank M.P. Encinar for fruitful discussions and comments on the content of the manuscript.    

\section*{Declaration of interest}

The authors report no conflict of interest.

\appendix
\section{Code validations: oscillations of a viscous drop in a quiescent fluid}

We validate the Cahn--Hilliard--Navier--Stokes code by simulating
the oscillations towards equilibrium of a drop in a quiescent fluid. 
We introduce a drop in a flow without velocity fluctuations,
and deform it slightly with an 
axisymmetric perturbation, so that its evolution towards equilibrium 
is linear and can be described in terms of axisymmetric spherical harmonics.
The solution of this flow was given by \citet{lamb1924hydrodynamics}, and reads 
\begin{equation}
r(\theta,t)=r_0 + \epsilon P_n(\cos\theta)\cos(f_n t)\exp(-\gamma_n t),
\end{equation}
where $r_0=d/2$ is the radius of the spherical drop, $\epsilon\ll r_0$ is the perturbation magnitude, and $P_n$ is the Legendre polynomial of order $n$.
Here $f_n$ and $\gamma_n$ are the characteristic oscillation frequency and decay rate of each harmonic,
\begin{equation}
f^2_n=\frac{\sigma}{\rho d^3}\frac{n(n-1)(n+1)(n+2)}{2n+1},
\end{equation}
and
\begin{equation}
\gamma_n=\frac{\nu}{d^2}\big(n(n-1) + (n+1)(n+2)\big),
\end{equation}
respectively. Here  $\sigma$ is the surface tension, $\rho$ and $\rho$ and $\nu$ the density and the kinematic viscosity of the fluid,
respectively. The two characteristic time scales of the problem are $\sqrt{\rho d^3/\sigma}$ and $d^2/\nu$. 
Their ratio, the Ohnesorge number, $\mathrm{\it Oh}=\sqrt{\rho/(\sigma d)}\nu$, is the only parameter of the problem.

\begin{figure}
\centering
\includegraphics[width=0.55\textwidth]{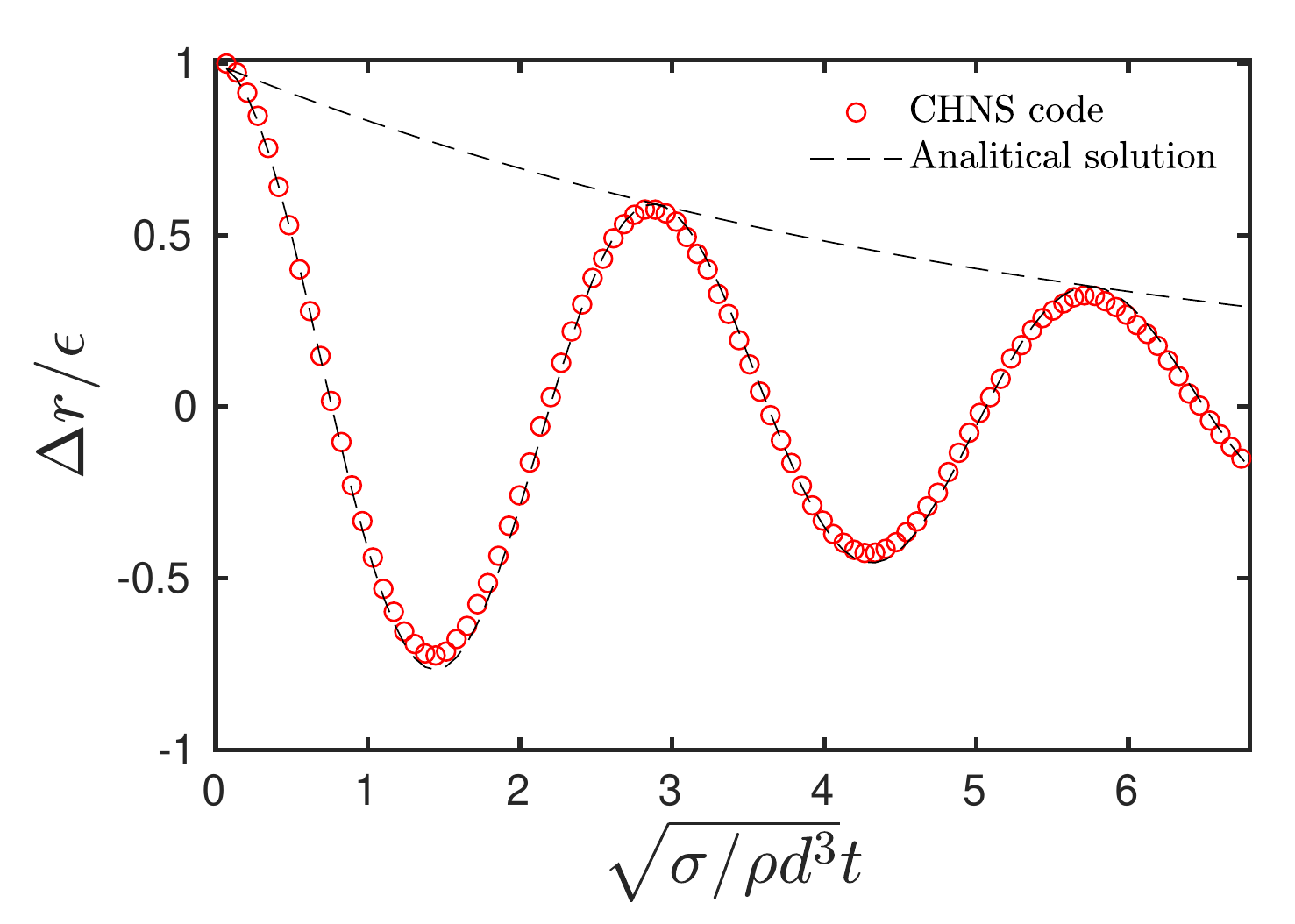}
% \vspace{2pt}\includegraphics[width=0.48\textwidth]{spectra_prod.pdf}
\caption{Comparison of the analytical solution of $\Delta r/\epsilon$ for a drop with $\mathrm{\it Oh}=0.013$ 
         with a numerical simulation with a Cahn-Hilliard-Navier-Stokes (CHNS) code.  
         } 
\label{validation}
\end{figure}

We consider a spherical drop with $\mathrm{\it Oh}=0.013$, and perturb only its second harmonic, $n=2$.
We measure the oscillation of the interface in the direction of the longest semiaxis of the drop ($\theta=0$ and $P_2=1$),
and compare it with the analytical solution.  The numerical parameters of the code are the same as those used in the single drop simulations.
In figure \ref{validation}, we show the evolution of 
\begin{equation}
{\Delta r}={r(0,t) - r_0},
\end{equation}
in the simulation, and the analytical solution,
\begin{equation}
{\Delta r}=\epsilon\cos(f_n t)\exp(-\gamma_n t).
\end{equation}
% In figure \ref{validation}, we compare the analytical solution of $\Delta r/\epsilon$ with the code results. a drop with $Oh=0.013$.
The match between the two is very good. The characteristic oscillatory frequency given by 
the code matches the analytical solution with less than $1\%$ error, whereas the decay rate is reproduced with less than $4\%$ error.
% We have checked that these results are similar for different $Oh$ in which the evolution of the shows an damped oscillations. 

\bibliographystyle{jfm}
\bibliography{draft}

\begin{thebibliography}{49}
\expandafter\ifx\csname natexlab\endcsname\relax\def\natexlab#1{#1}\fi
\def\au#1{#1} \def\ed#1{#1} \def\yr#1{#1}\def\at#1{#1}\def\jt#1{\textit{#1}}
  \def\bt#1{#1}\def\bvol#1{\textbf{#1}} \def\vol#1{#1} \def\pg#1{#1}
  \def\publ#1{#1}\def\arxiv#1{#1}\def\org#1{#1}\def\st#1{\textit{#1}}

\bibitem[Aiyer {\em et~al.\/}(2019)Aiyer, Yang, Chamecki \&
  Meneveau]{aiyer2019population}
{\sc \au{Aiyer, A.K.}, \au{Yang, D.}, \au{Chamecki, M.} \& \au{Meneveau, C.}}
  \yr{2019}  \at{A population balance model for large eddy simulation of
  polydisperse droplet evolution}.  \jt{J.\ Fluid Mech.}  \bvol{878},
  \pg{700--739}.

\bibitem[Andersson \& Andersson(2006)]{andersson2006}
{\sc \au{Andersson, R.} \& \au{Andersson, B.}} \yr{2006}  \at{On the breakup of
  fluid particles in turbulent flows}.  \jt{AIChE Journal}  \bvol{52},
  \pg{2020--2030}.

\bibitem[Ashurst {\em et~al.\/}(1987)Ashurst, Kerstein, Kerr \&
  Gibson]{ashurst1987alignment}
{\sc \au{Ashurst, W.T.}, \au{Kerstein, A.R.}, \au{Kerr, R.M.} \& \au{Gibson,
  C.H.}} \yr{1987}  \at{Alignment of vorticity and scalar gradient with strain
  rate in simulated navier--stokes turbulence}.  \jt{Phys. Fluids}  \bvol{30},
  \pg{2343--2353}.

\bibitem[Badalassi {\em et~al.\/}(2003)Badalassi, Ceniceros \&
  Banerjee]{badalassi2003computation}
{\sc \au{Badalassi, V.E.}, \au{Ceniceros, H.D.} \& \au{Banerjee, S.}} \yr{2003}
   \at{Computation of multiphase systems with phase field models}.  \jt{J.\
  Comput.\ Phys.}  \bvol{190},  \pg{371--397}.

\bibitem[Betchov(1956)]{betchov1956inequality}
{\sc \au{Betchov, R}} \yr{1956}  \at{An inequality concerning the production of
  vorticity in isotropic turbulence}.  \jt{J.\ Fluid Mech.}  \bvol{1}~(5),
  \pg{497--504}.

\bibitem[Buaria {\em et~al.\/}(2020)Buaria, Bodenschatz \&
  Pumir]{buaria2020vortex}
{\sc \au{Buaria, D.}, \au{Bodenschatz, E.} \& \au{Pumir, A.}} \yr{2020}
  \at{Vortex stretching and enstrophy production in high reynolds number
  turbulence}.  \jt{Phys. Rev. Fluids}  \bvol{5}~(10),  \pg{104602}.

\bibitem[Calabrese {\em et~al.\/}(1986)Calabrese, Chang \&
  Dang]{calabrese1986drop}
{\sc \au{Calabrese, R.V.}, \au{Chang, T.P.K.} \& \au{Dang, P.T.}} \yr{1986}
  \at{Drop breakup in turbulent stirred-tank contactors. {P}art {I}: {E}ffect
  of dispersed-phase viscosity}.  \jt{AIChE J.}  \bvol{32},  \pg{657--666}.

\bibitem[Cardesa {\em et~al.\/}(2017)Cardesa, Vela-Mart{\'\i}n \&
  Jim{\'e}nez]{cardesa2017}
{\sc \au{Cardesa, J.I.}, \au{Vela-Mart{\'\i}n, A.} \& \au{Jim{\'e}nez, J.}}
  \yr{2017}  \at{The turbulent cascade in five dimensions}.  \jt{Science}
  \bvol{357},  \pg{782--784}.

\bibitem[Chen \& Shen(1998)]{chen1998applications}
{\sc \au{Chen, L.Q.} \& \au{Shen, J.}} \yr{1998}  \at{Applications of
  semi-implicit fourier-spectral method to phase field equations}.
  \jt{Comput.\ Phys.\ Commun.}  \bvol{108},  \pg{147--158}.

\bibitem[Coulaloglou \& Tavlarides(1976)]{coulaloglou1976drop}
{\sc \au{Coulaloglou, C.A.} \& \au{Tavlarides, L.L.}} \yr{1976}  \at{Drop size
  distributions and coalescence frequencies of liquid-liquid dispersions in
  flow vessels}.  \jt{AIChE} ~(2),  \pg{289--297}.

\bibitem[Dodd \& Ferrante(2016)]{dodd2016interaction}
{\sc \au{Dodd, M.S.} \& \au{Ferrante, A.}} \yr{2016}  \at{On the interaction of
  taylor length scale size droplets and isotropic turbulence}.  \jt{J. Fluid
  Mech.}  \bvol{806},  \pg{356--412}.

\bibitem[Dopazo {\em et~al.\/}(2000)Dopazo, Lozano \&
  Barreras]{dopazo2000vorticity}
{\sc \au{Dopazo, C.}, \au{Lozano, A.} \& \au{Barreras, F.}} \yr{2000}
  \at{Vorticity constraints on a fluid/fluid interface}.  \jt{Phys. Fluids}
  \bvol{12},  \pg{1928--1931}.

\bibitem[Eastwood {\em et~al.\/}(2004)Eastwood, Armi \& Lasheras]{eastwood2004}
{\sc \au{Eastwood, C.D.}, \au{Armi, L.} \& \au{Lasheras, J.C.}} \yr{2004}
  \at{The breakup of immiscible fluids in turbulent flows}.  \jt{J.\ Fluid
  Mech.}  \bvol{502},  \pg{309–333}.

\bibitem[Elghobashi(2019)]{elghobashi2019}
{\sc \au{Elghobashi, S}} \yr{2019}  \at{Direct numerical simulation of
  turbulent flows laden with droplets or bubbles}.  \jt{Annu.\ Rev.\ Fluid
  Mech.} ~(0).

\bibitem[Girimaji \& Pope(1990)]{girimaji1990material}
{\sc \au{Girimaji, S.S.} \& \au{Pope, S.B.}} \yr{1990}  \at{Material-element
  deformation in isotropic turbulence}.  \jt{J. Fluid Mech.}  \bvol{220},
  \pg{427--458}.

\bibitem[Hakansson(2019)]{hakansson2019}
{\sc \au{Hakansson, A.}} \yr{2019}  \at{Emulsion formation by homogenization:
  Current understanding and future perspectives}.  \jt{Annu.\ Rev.\ Food Sci.\
  Technol.}  \bvol{10},  \pg{239--258}.

\bibitem[Hakansson(2020)]{haakansson2020validity}
{\sc \au{Hakansson, A.}} \yr{2020}  \at{On the validity of different methods to
  estimate breakup frequency from single drop experiments}.  \jt{Chem. Engin.
  Sci.}  \bvol{227},  \pg{115908}.

\bibitem[Hamlington {\em et~al.\/}(2008)Hamlington, Schumacher \&
  Dahm]{hamlington2008local}
{\sc \au{Hamlington, P.E.}, \au{Schumacher, J.} \& \au{Dahm, W.J.A.}} \yr{2008}
   \at{Local and nonlocal strain rate fields and vorticity alignment in
  turbulent flows}.  \jt{Phys. Rev. E}  \bvol{77}~(2),  \pg{026303}.

\bibitem[Hinze(1955)]{hinze1955}
{\sc \au{Hinze, J.O.}} \yr{1955}  \at{Fundamentals of the hydrodynamic
  mechanism of splitting in dispersion processes}.  \jt{AIChE Journal}
  \bvol{1}~(3),  \pg{289--295}.

\bibitem[Jacqmin(1999)]{jacqmin1999calculation}
{\sc \au{Jacqmin, D.}} \yr{1999}  \at{Calculation of two-phase
  {N}avier--{S}tokes flows using phase-field modeling}.  \jt{J.\ Comput.\
  Phys.}  \bvol{155},  \pg{96--127}.

\bibitem[Jim{\'e}nez {\em et~al.\/}(1993)Jim{\'e}nez, Wray, Saffman \&
  Rogallo]{jimenez1993structure}
{\sc \au{Jim{\'e}nez, J.}, \au{Wray, A.}, \au{Saffman, P.G.} \& \au{Rogallo,
  R.S.}} \yr{1993}  \at{The structure of intense vorticity in isotropic
  turbulence}.  \jt{J. Fluid Mech.}  \bvol{255},  \pg{65--90}.

\bibitem[Kolmogorov(1949)]{kolmogorov1949}
{\sc \au{Kolmogorov, A.N.}} \yr{1949} On the disintegration of drops in
  turbulent flow.  \bt{In {\em Doklady Akad. Nauk. USSR\/}}, ,  \vol{vol.~66},
  \pg{p. 825}.

\bibitem[Komrakova {\em et~al.\/}(2015)Komrakova, Eskin \&
  Derksen]{komrakova2015numerical}
{\sc \au{Komrakova, Alexandra~E}, \au{Eskin, Dmitry} \& \au{Derksen, JJ}}
  \yr{2015}  \at{Numerical study of turbulent liquid-liquid dispersions}.
  \jt{AIChE Journal}  \bvol{61}~(8),  \pg{2618--2633}.

\bibitem[Lamb(1924)]{lamb1924hydrodynamics}
{\sc \au{Lamb, H.}} \yr{1924} {\em Hydrodynamics\/}.  \publ{University Press}.

\bibitem[Lasheras {\em et~al.\/}(2002)Lasheras, Eastwood, Mart{\i}nez-Baz{\'a}n
  \& Montanes]{lasheras2002review}
{\sc \au{Lasheras, J.C.}, \au{Eastwood, C.}, \au{Mart{\i}nez-Baz{\'a}n, C.} \&
  \au{Montanes, J.L.}} \yr{2002}  \at{A review of statistical models for the
  break-up of an immiscible fluid immersed into a fully developed turbulent
  flow}.  \jt{Int. J. Multiph. Flow}  \bvol{28},  \pg{247--278}.

\bibitem[Liao \& Lucas(2009)]{liao2009literature}
{\sc \au{Liao, Y.} \& \au{Lucas, D.}} \yr{2009}  \at{A literature review of
  theoretical models for drop and bubble breakup in turbulent dispersions}.
  \jt{Chem.\ Eng.\ Sci.}  \bvol{64},  \pg{3389--3406}.

\bibitem[Luo \& Svendsen(1996)]{luo1996}
{\sc \au{Luo, H.} \& \au{Svendsen, H.}} \yr{1996}  \at{Theoretical model for
  drop and bubble breakup in turbulent dispersions}.  \jt{AIChE J.}  \bvol{42},
   \pg{1225--1233}.

\bibitem[Maa\ss\ \& Kraume(2012)]{maass2012}
{\sc \au{Maa\ss\, S.} \& \au{Kraume, M.}} \yr{2012}  \at{Determination of
  breakage rates using single drop experiments}.  \jt{Chem.\ Eng.\ Sci.}
  \bvol{70},  \pg{146 -- 164}.

\bibitem[Magaletti {\em et~al.\/}(2013)Magaletti, Picano, Chinappi, Marino \&
  Casciola]{magaletti2013sharp}
{\sc \au{Magaletti, F.}, \au{Picano, F.}, \au{Chinappi, M.}, \au{Marino, L.} \&
  \au{Casciola, C.~M.}} \yr{2013}  \at{The sharp-interface limit of the
  {C}ahn--{H}illiard/{N}avier--{S}tokes model for binary fluids}.  \jt{J. Fluid
  Mech.}  \bvol{714},  \pg{95--126}.

\bibitem[Mukherjee {\em et~al.\/}(2019)Mukherjee, Safdari, Shardt,
  Kenjere{\v{s}} \& Van~den Akker]{mukherjee2019droplet}
{\sc \au{Mukherjee, S.}, \au{Safdari, A.}, \au{Shardt, O.}, \au{Kenjere{\v{s}},
  S.} \& \au{Van~den Akker, H.E.A.}} \yr{2019}  \at{Droplet--turbulence
  interactions and quasi-equilibrium dynamics in turbulent emulsions}.  \jt{J.
  Fluid Mech.}  \bvol{878},  \pg{221--276}.

\bibitem[Narsimhan {\em et~al.\/}(1979)Narsimhan, Gupta \&
  Ramkrishna]{narsimhan1979model}
{\sc \au{Narsimhan, G.}, \au{Gupta, J.P.} \& \au{Ramkrishna, D.}} \yr{1979}
  \at{A model for transitional breakage probability of droplets in agitated
  lean liquid-liquid dispersions}.  \jt{Chem. Eng. Sci.} ~(2),  \pg{257--265}.

\bibitem[Ohkitani \& Kishiba(1995)]{ohkitani1995nonlocal}
{\sc \au{Ohkitani, K.} \& \au{Kishiba, S.}} \yr{1995}  \at{Nonlocal nature of
  vortex stretching in an inviscid fluid}.  \jt{Phys. Fluids}  \bvol{7},
  \pg{411--421}.

\bibitem[Perlekar(2019)]{perlekar2019kinetic}
{\sc \au{Perlekar, P.}} \yr{2019}  \at{Kinetic energy spectra and flux in
  turbulent phase-separating symmetric binary-fluid mixtures}.  \jt{J. Fluid
  Mech.}  \pg{pp. 459--474}.

\bibitem[Perlekar {\em et~al.\/}(2012)Perlekar, Biferale, Sbragaglia,
  Srivastava \& Toschi]{perlekar2012droplet}
{\sc \au{Perlekar, P.}, \au{Biferale, L.}, \au{Sbragaglia, M.}, \au{Srivastava,
  S.} \& \au{Toschi, F.}} \yr{2012}  \at{Droplet size distribution in
  homogeneous isotropic turbulence}.  \jt{Phys. Fluids} ~(6),  \pg{065101}.

\bibitem[Qian {\em et~al.\/}(2006)Qian, McLaughlin, Sankaranarayanan,
  Sundaresan \& Kontomaris]{qian2006simulation}
{\sc \au{Qian, D.}, \au{McLaughlin, J.~B.}, \au{Sankaranarayanan, K.},
  \au{Sundaresan, S.} \& \au{Kontomaris, K.}} \yr{2006}  \at{Simulation of
  bubble breakup dynamics in homogeneous turbulence}.  \jt{Chem. Eng. Commun.}
  \bvol{193}~(8),  \pg{1038--1063}.

\bibitem[Rallison(1984)]{rallison1984}
{\sc \au{Rallison, J.M.}} \yr{1984}  \at{The deformation of small viscous drops
  and bubbles in shear flows}.  \jt{Annu.\ Rev.\ Fluid Mech.}  \bvol{16},
  \pg{45--66}.

\bibitem[Ramkrishna \& Singh(2014)]{ramkrishna2014population}
{\sc \au{Ramkrishna, D.} \& \au{Singh, M.R.}} \yr{2014}  \at{Population balance
  modeling: current status and future prospects}.  \jt{Annu. Rev. Chem. Biomol.
  Eng.}  \pg{pp. 123--146}.

\bibitem[Risso \& Fabre(1998)]{risso1998}
{\sc \au{Risso, F.} \& \au{Fabre, J.}} \yr{1998}  \at{Oscillations and breakup
  of a bubble immersed in a turbulent field}.  \jt{J. Fluid Mech.}  \bvol{372},
   \pg{323--355}.

\bibitem[Roccon {\em et~al.\/}(2017)Roccon, De~Paoli, Zonta \&
  Soldati]{roccon2017viscosity}
{\sc \au{Roccon, A.}, \au{De~Paoli, M.}, \au{Zonta, F.} \& \au{Soldati, A.}}
  \yr{2017}  \at{Viscosity-modulated breakup and coalescence of large drops in
  bounded turbulence}.  \jt{Phys. Rev. Fluids}  \bvol{2}~(8),  \pg{083603}.

\bibitem[Rosti {\em et~al.\/}(2019)Rosti, Ge, Jain, Dodd \&
  Brandt]{rosti2019droplets}
{\sc \au{Rosti, M.E.}, \au{Ge, Z.}, \au{Jain, S.S.}, \au{Dodd, M.S.} \&
  \au{Brandt, L.}} \yr{2019}  \at{Droplets in homogeneous shear turbulence}.
  \jt{J. of Fluid Mech.}  \pg{pp. 962--984}.

\bibitem[Scarbolo {\em et~al.\/}(2015)Scarbolo, Bianco \&
  Soldati]{scarbolo2015coalescence}
{\sc \au{Scarbolo, L.}, \au{Bianco, F.} \& \au{Soldati, A.}} \yr{2015}
  \at{Coalescence and breakup of large droplets in turbulent channel flow}.
  \jt{Phys. Fluids}  \bvol{27},  \pg{073302}.

\bibitem[Shao {\em et~al.\/}(2018)Shao, Luo, Yang \& Fan]{shao2018}
{\sc \au{Shao, C.}, \au{Luo, K.}, \au{Yang, Y.} \& \au{Fan, J.}} \yr{2018}
  \at{Direct numerical simulation of droplet breakup in homogeneous isotropic
  turbulence: The effect of the weber number}.  \jt{Int.\ J.\ Multiph.\ Flow}
  \bvol{107},  \pg{263--274}.

\bibitem[Soligo {\em et~al.\/}(2019)Soligo, Roccon \& Soldati]{soligo2019mass}
{\sc \au{Soligo, G.}, \au{Roccon, A.} \& \au{Soldati, A.}} \yr{2019}
  \at{Mass-conservation-improved phase field methods for turbulent multiphase
  flow simulation}.  \jt{Acta Mechanica}  \bvol{230},  \pg{683--696}.

\bibitem[Soligo {\em et~al.\/}(2020)Soligo, Roccon \&
  Soldati]{soligo2020effect}
{\sc \au{Soligo, G.}, \au{Roccon, A.} \& \au{Soldati, A.}} \yr{2020}
  \at{Effect of surfactant-laden droplets on turbulent flow topology}.
  \jt{Phys. Rev. Fluids}  \bvol{5},  \pg{073606}.

\bibitem[Solsvik \& Jakobsen(2015)]{solsvik2015single}
{\sc \au{Solsvik, J.} \& \au{Jakobsen, H.A.}} \yr{2015}  \at{Single drop
  breakup experiments in stirred liquid--liquid tank}.  \jt{Chem. Eng. Sci.}
  \bvol{131},  \pg{219--234}.

\bibitem[Yi {\em et~al.\/}(2021)Yi, Toschi \& Sun]{yi2021global}
{\sc \au{Yi, Lei}, \au{Toschi, Federico} \& \au{Sun, Chao}} \yr{2021}
  \at{Global and local statistics in turbulent emulsions}.  \jt{Journal of
  Fluid Mechanics}  \bvol{912}.

\bibitem[Yue {\em et~al.\/}(2007)Yue, Zhou \& Feng]{Yue2007}
{\sc \au{Yue, P.}, \au{Zhou, C.} \& \au{Feng, J.~J.}} \yr{2007}
  \at{Spontaneous shrinkage of drops and mass conservation in phase{-}field
  simulations}.  \jt{J.\ Comput.\ Phys.}  \bvol{223},  \pg{1--9}.

\bibitem[Zhang \& Ye(2017)]{zhang2017flux}
{\sc \au{Zhang, Yujie} \& \au{Ye, Wenjing}} \yr{2017}  \at{A flux-corrected
  phase-field method for surface diffusion}.  \jt{Commun. Comput.\ Phys.}
  \bvol{22},  \pg{422--440}.

\bibitem[Zhong \& Komrakova(2019)]{zhong2019liquid}
{\sc \au{Zhong, Cheng} \& \au{Komrakova, Alexandra}} \yr{2019}  \at{Liquid drop
  breakup in homogeneous isotropic turbulence}.  \jt{Int. J. Numer. Method H.}
  .

\end{thebibliography}

% \begin{itemize}
% 
% The geometry of the velocity gradients close to the surface is quasi-two d, with a typical. $-6:1:5$ at $We=50$

\end{document}